\documentclass[mnsc,nonblindrev]{informs3}

\OneAndAHalfSpacedXI 

\usepackage{natbib}
 \bibpunct[, ]{(}{)}{,}{a}{}{,}%
 \def\bibfont{\small}%
\TheoremsNumberedThrough     
\ECRepeatTheorems

\pdfoutput=1

\usepackage{amsmath}
\usepackage{cases}
\usepackage{algorithm}
\usepackage{mathrsfs}
\usepackage{algpseudocode}
\usepackage{makecell}
\usepackage{tabularx}
\usepackage{tikz}
\usepackage{tikz-3dplot}
\usetikzlibrary{arrows.meta,intersections,calc}
\usepackage{graphicx}
\usepackage{subfigure}
\usepackage{enumerate}

\usepackage[]{hyperref}

\allowdisplaybreaks[2]

\MANUSCRIPTNO{MS-OPM-2023-03005}

\makeatletter
\newenvironment{breakablealgorithm}
  {
   \begin{center}
     \refstepcounter{algorithm}
     \hrule height.8pt depth0pt \kern2pt
     \renewcommand{\caption}[2][\relax]{
       {\raggedright\textbf{\ALG@name~\thealgorithm} ##2\par}%
       \ifx\relax##1\relax 
         \addcontentsline{loa}{algorithm}{\protect\numberline{\thealgorithm}##2}%
       \else 
         \addcontentsline{loa}{algorithm}{\protect\numberline{\thealgorithm}##1}%
       \fi
       \kern2pt\hrule\kern2pt
     }
  }{
     \kern2pt\hrule\relax
   \end{center}
  }
\makeatother

\begin{document}

\RUNAUTHOR{Liu et.al}

\RUNTITLE{Sewage Discharging in a Line}


\TITLE{Sewage Discharging in a Line: \\Global Optimization and Grand Cooperation}


\ARTICLEAUTHORS{%
\AUTHOR{Xucheng Liu}
\AFF{
International Institute of Finance, School of Management, University of Science and Technology of China\\ \EMAIL{lkdn@mail.ustc.edu.cn}}
\AUTHOR{Lindong Liu}
\AFF{
International Institute of Finance, School of Management, University of Science and Technology of China\\ \EMAIL{njulld@gmail.com}}
\AUTHOR{Yifu Li}
\AFF{
International Institute of Finance, School of Management, University of Science and Technology of China\\ \EMAIL{yifuli@ustc.edu.cn}}
\AUTHOR{Anran Li}
\AFF{Department of Decision Sciences and Managerial Economics, Business School, The Chinese University of Hong Kong\\ 
\EMAIL{anranli@cuhk.edu.hk}}
} 

\ABSTRACT{%
Players cooperating in a line is a special while essential phenomenon in real life collaborating activities such as assembly line production, pipeline supply chain management and other streamlining operational settings.
In this paper, we study the scenario of cooperative sewage discharge with multiple participants positioning in a line along a river such that the optimization decision and cooperation strategy are mutually affected by both upstream and downstream players.
We make three main contributions accordingly: Firstly, we formalize the sewage discharge problem (SDP) for different groups of players, and use greedy strategy and dynamic programming to design the optimal algorithms to solve the SDP in polynomial time.
Secondly, we show that the cooperative game defined on sewage discharge problem, referred to as SDG, has a non-empty core due to its special line-positioning structure. Therefore, a grand stable cooperation is guaranteed.
Furthermore, inspired by the fact that the SDG is core non-empty while non-convex, we successfully identify a relaxed concept of convexity--directional-convexity, which can also serve as a sufficient condition for a cooperative game having a non-empty core.
}




\maketitle


\section{Introduction}
Rivers, known as birthplaces of many ancient cultures, have served as essential resources for various agricultural, domestic and industrial purposes. 
Take the Colorado River Basin as an example, it serves more than 40 million  people in the southwestern United States and northwestern Mexico \citep{wheeler2022will}. Tremendous wealth has been created in agriculture, municipal and industrial fields by the river.
The river basin is a complex and crucial system, that affects both economic development and the daily life of inhabitants.
Entities share rivers and earn profits through activities such as food production, transportation, power generation, and sewage discharge \citep[see,][]{just1998conflict,kramer2002estimating,cai2003physical,chen2012socio,james2014economic,galelli2022opportunities}. 
Therefore, river-sharing management plays a vital role in achieving sustainable development, and sustaining regional stable and balance.

As public resources, rivers are not owned by any entity, but they can be used by all entities freely. This makes the river-sharing problem very difficult. When sharing some classical resources, such as in bin packing problems \citep{liu2009complexity}, newsvendor problems \citep{chen2019population} and machine scheduling problems \citep{schulz2013approximating,liu2018simultaneous}, the property rights over these resources are well determined, and these resources could be conveniently reallocated among entities to achieve global optimization.
However, the sharing of rivers often leads to numerous disputes, particularly due to the flow direction that dividing territories into upstream and downstream regions. In this context, rights of the downstream entity to access the river are contingent upon the river usage by the upstream entity. Similar scenarios can be observed in various other contexts such as supply chain management and electricity transmission.
Up to now, disputes over river-sharing continue to arise.\footnote{For instance, the International Court of Justice made the judgement on $1^{\text{st}}$ Dec. 2022, in the \emph{Dispute over the Status and Use of the Waters of the Silala (Chile v. Bolivia)} case, the international status of the rivers in question by customary rules of international law, https://www.icj-cij.org/en/case/162/press-releases.} Without pre-agreed economic compensation agreement between the upstream and downstream, the downstream may think that the upstream is overusing the river and negatively affecting its own economy. To this end, when resolving the river-sharing problems, the principal should make a centralized decision considering both the upstream and downstream entities, and treat it as a global optimization problem.

Such a river-sharing optimization problem can be modeled as a cooperative game, where multiple entities participate and form a grand coalition to pursue global optimization, and the third party fairly divides the global benefits generated (or, cost incurred) among entities. As any subset of entities can talk and coalize, an inappropriate profit allocation will cause instability of the game. Therefore, we hope to seek for a stable grand coalition and a fair allocation of profit so no one has incentive to deviate. However, the constraints to be considered may be exponential in the number of players, which makes the problem rather difficult.

Regarding the studies that apply cooperative game theory to the river-sharing problem, the literature can be divided into two strands: water resource sharing utilizing the river \citep[e.g. ][]{demange2004on,ansink2012sequential,dawande2013efficient} and treatment cost sharing cleaning the river \citep[e.g. ][]{rua2013sharing,brink2017polluted,steinmann2019sharing}.
In the first strand, every player benefits from the utilization and redistribution of the water resource in the river cooperatively.
In the second strand, the responsibility of cleaning the polluted river is shared among the players cooperatively.
To our knowledge that all the existing studies are devoted to proposing economically efficient allocation schemes of the global benefits (or cost) for resolving river disputes while making little use of the distinct cooperative structure of entities for the purpose of guaranteeing the stable cooperation, which is crucial to the social sustainable development.

Currently, sustainable operations management has attracted an increasing impact on the OM/OR 
domain \citep{atasu2020sustainable}. Government and international agencies have begun to embrace the broad concept of sustainable development, less pollution and more benefits \citep{angell1999integrating,drake2013om}, among which sewage discharging management is an important part \citep{hammond2021detection}.
In this paper, we investigate the cooperative sewage discharge problem in the river sharing.
To be specific, we consider the situation where multiple entities, say firms, are located in a line along a river and discharging sewage arising from their economic activities.
The sewage discharging quota of each entity is regulated by various practical concerns such as its production capacity and river water quality restrictions.
A third party, the social planner such as a local government, is interested in achieving regional sustainable development via global optimization such that economic benefits are generated without deteriorating the river water quality. The foundation of such sustainable operations is a fair benefit allocation scheme that guarantees stable cooperation among the entities.
Therefore, to incentive ever-lasting cooperation for sustainable operation management, the following crucial questions ought to be answered. 
(i) What is the sewage discharging quota of each entity if we want to optimize the group benefits?
(ii) Is it possible to fairly allocate the benefit generated from economic activities such that the grand coalition is stable?
(iii) As the entities have a line-positioning structure introduced by the river, the question is: How does the line-positioning structure affect the cooperation stability of the entities?

Based on the cooperative game theory and optimization methods, we build up a mathematical model to answer the above questions.
First, we model the sewage discharge problem as a nonlinear program with the objective of maximizing total production profit while subject to a set of pollution limits. We derive the optimal sewage discharging scheme under various conditions.
While the first model is under control of a central authority who has control over all firms with the objective of maximizing the total profit, we care more about the second question---how would the firms behave if they have own jurisdiction? Would they form a sub-coalition? Would the grand coalition be stable? To answer these questions, we axiomatize the positional advantage of each entity and investigate the coalition formation mechanism with a cooperation measurement method. Building on these, we discuss the properties of the characteristic function of the sewage discharge game. Interestingly, we show non-emptiness of core under this sewage discharge game and derive an allocation that guarantees stability of the grand coalition.
Third, we extend our analysis to a relaxed concept of convexity, referred to as directional-convexity, which involves the line-positioning structure of entities. We derive a class of core allocations of the directional-convex game, and show that directional-convexity is a sufficient condition for the balanced cooperative game. 

To our knowledge, our paper is the first to address the grand cooperation of the cooperative sewage discharge problem for sustainable operations management. We contribute to the literature through the following three aspects:

First, we recognize that the inherent self-purification capacity of the river and distinguishing production capacities of entities forming the foundation for cooperative sewage discharge practices. Therefore, the central authority may want to rearrange the sewage discharging quota of each entity to obtain the maximum benefits. To achieve this, we develop two polynomial time-solvable algorithms, which can compute the socially optimal sewage discharge scheme efficiently.

Second, built on the Bondareva-Shapley Theorem, we prove that the sewage discharge game (SDG) is balanced. This indicates that the central authority can govern entities to achieve sustainable economic growth under proper pollution control.
For the case where the river goes through multiple jurisdictions, a balanced SDG also indicates that there is at least one profit allocation scheme that can stabilize the grand coalition and resolve regional dispute. Our study provides the theoretical foundation to resolve the river-resource dispute in practice.

Third, as a sufficient condition for the balanced cooperative game, directional-convexity could be a screening tool for the stable grand coalition in OR/OM problems. In directional-convex games, the central authority can flexibly use convex combination of core allocations that we proposed, to minimize the dissatisfactions in the grand coalition. Our results provide theoretical foundation for cooperation among countries which is the foundation of regional sustainable development.

The rest of this paper is organized as follows: 
In Section \ref{lr} we review the related literature. 
In Section \ref{sec2}, we discuss some preliminaries about the river problem and cooperative game. In Section \ref{sec3}, we study the sewage discharge problem and introduce an exact algorithm to compute the optimal solution. In Section \ref{sec4}, we introduce the sewage discharge game, and prove the non-emptiness of the core and provide a core allocation. We extend the discussion to the directional-convex games in Section \ref{sec5}. 
Finally, we concluded the paper in Section \ref{sec6}.

\section{Literature Review}\label{lr}

In this paper, we consider a cooperative sewage discharge problem in which multiple entities along a river jointly decide the sewage discharge and group benefit allocation. We contribute to the literature along three streams: river administration problems, river-sharing games, and sustainable operations management.

\subsection{River Administration Problems}

River administration problem is a long-standing but ever flourishing research topic investigated by scholars from various domains. \cite{simonovic2012managing} provides an overview of water resource systems management in this area. 
Some of the most prominent examples related to river administration problems include river water trade \citep[e.g.,][]{weber2001markets,wang2011trading}, irrigation water management \citep[e.g.,][]{marques2005modeling,dawande2013efficient}, reservoir location \citep[e.g.,][]{mun2021designing}, to name just a few.

In this paper, we focus on the topic of sewage discharge problems, which is closely related to the literature on river pollution control problems. The main interest in studying this domain is monitoring and regulating sewage pollution in water bodies. For instance, \cite{fiacco1982sensitivity} introduce a nonlinear water pollution control model and conduct a sensitivity analysis on the optimal waste treatment cost using a dataset of the upper Hudson River. \cite{meyer2019real} present an approach to enable real-time monitoring of various river parameters, and identify different pollution situations in a river. Recently, \cite{hammond2021detection} use machine learning methods to detect untreated wastewater discharges, which helps companies identify malfunctioning water treatment equipment and the government detect unregulated discharge facilities. \cite{muthulingam2022does} introduce an econometric method to examine the impact of water scarcity on toxic water release of the factories in Texas.

Our work also considers sewage mixing and pollution natural degradation. However, we study the problem under the cooperative game framework rather than by examining the dataset empirically. We formulate the sewage discharge problem as an optimization problem and investigate how to improve social welfare considering the sewage discharging demand and pollution control constraints. Our study offers a new method to investigate water administration problems. 

\subsection{River-Sharing Games}

In this paper, we build up a game theory model to investigate the river-sharing problems, which belongs to the area of river-sharing games. The river-sharing game usually deals with the allocation of the benefits (or costs) among a set of entities sharing the river, see \cite{beal2013the} for a broad overview of the theory and the practice of river-sharing games.

The study of river-sharing game can be traced to the work of \cite{suzuki1976cost}. They show that, in exploiting the water resource, cooperative game theory could offer guidelines in assigning fair benefits to the entities in a cooperative venture.
Then, a stylized model for river-sharing problem is introduced by \cite{ambec2002sharing}, who is the first to study how shall the countries or cities along a river share the water resources by formulating a cooperative game. They propose a downstream incremental distribution which is the unique distribution satisfying the core lower bounds and the aspiration upper bounds.
Later, \cite{ambec2008sharing} extend the discussion to the river-sharing model with satiable entities, and individual rationality constraints.
They reveal that the downstream incremental distribution satisfies all core lower bounds for all the connected coalitions. 

Besides investigating how to share the water resources, \cite{ni2007sharing} study the sharing problem of a polluted river. In order to use the water resource in the river, the entities must take responsibility for cleaning the river. To share the pollutant cleaning costs, they propose the Local Responsibility Sharing method and the Upstream Equal Sharing method which coincide with the Shapley value \citep{shapley1953value}. They show that these two methods are in the cores of the corresponding games.
Recently, \cite{dong2012sharing} study a general scenario of sharing polluted rivers with multiple springs.
They propose the Downstream Equal Sharing method and show that the optimal value obtained by this method equals the Shapley value of specific games in characteristic form.
\cite{alcalde2015sharing} propose the Upstream Responsibility Rule considering the uncertainty of the pollutant transfer rate.

In our paper, we investigate the properties which affect the grand coalition stability of the sewage discharge game. We focus more on the distinct cooperative structure of the entities and analyze the stability of cooperation from the perspective of the core definition. By the duality theory, we prove the existence of a solution to the dual problem of the core allocation of our sewage discharge game, which indicates the stability of the grand coalition formed by all entities along the river. In addition, we axiomatize a sufficient condition that guarantees the core non-emptiness of sewage discharge games and extend it to a class of general balanced cooperative games.
 
\subsection{Sustainable Operations Management}

Our work also belongs to the area of sustainable operations management. \cite{williams2017systems} provide a summary on sustainable operations management, and \cite{atasu2020sustainable} review the latest advances in this area. 

As sustainable operations management is a broad area with many streams, we briefly summarize the relevant literature. Our study is closely related to the literature on environment protection policy design, renewable resource exploitation, and emission control in the manufacturing process. 

On the environmental protection policy analysis, \cite{drake2013om} study the environmental protection practices and suggest that it is the key to sustainable operations management. \cite{murali2019effects} investigate the green product development problem with game theory. They formulate a consumer-driven model and show how government should stimulate sustainable operations management. 
With the development of technology, the exploitation of renewable resources has become one of the important environmental protection practices. Hence, some scholars study the investment and marketing decisions for renewable resources \citep[see][]{lohndorf2013optimizing,hu2015capacity,aflaki2017strategic}. 
Other researchers focus on investigating how to control emissions in the manufacturing process. For example, \cite{cachon2014retail} introduce a retail supply chain model. They find that improving fuel efficiency is a more effective carbon reduction mechanism than adopting carbon pricing policies.
\cite{sunar2016allocating} study the process emissions allocation in co-production and show that imposing the emission tax on the primary product can greatly reduce carbon emission.

We contribute literature by developing a coalition formation mechanism in the cooperative sewage discharge game. In river pollution issues such as the eutrophication of water bodies, our coalition formation mechanism can help allocate the emission quotas and allocate the pollution control responsibilities. We find that promoting cooperation among entities can increase the efficiency in the renewable resources such as river water, which lead to a ``less pollution and higher collective interests'' situation. 

\section{Preliminaries}\label{sec2}
To formally analyze the optimization problem and cooperation strategy of players in sewage discharging, for preparation, we now introduce some preliminaries and notations regarding to water quality model and the cooperative game theory in this section.

\subsection{Water Quality Model}\label{sec2.1}

Water quality model is widely used to describe the migration and transformation of water pollutants in water environmentology, and has been studied intensively within many research areas such as hydrology and hydrodynamics, limnology and chemistry.
While the relevant studies are comprehensive and important for practical implementations for water quality management, our focus in this paper is to investigate the integrated concept of cooperative sewage discharge game, and it is sufficient for us to refer to two classic water quality models proposed in \cite{chapra2008surface} to capture certain key features.

The first model, called the \emph{uniform mixture model}, is designed for evaluating the water quality of a river after it is mixed by sewage.
In general, this model omits the detailed river-sewage mixing process and simply assumes that the flow rate of the river is constant, and that the concentration of pollutants after river-sewage mixing would immediately reach a uniformly homogeneous state at the river cross-section of the sewage outlet.
In this case, the pollutant concentration value $C$ after river-sewage mixing satisfies $C(Q_r+Q_p)=C_rQ_r+C_pQ_p$, where $C_r$ and $C_p$ are the respective pollutant concentration values of the river and the sewage before mixing, and $Q_r$ and $Q_p$ are the respective flow rates of the river and the sewage.
Accordingly, at any river cross-section, the \emph{pollution level} denotes the mass flow rate of pollutants, which is the product of the pollutant concentration and flow rate. It is then clear that at any sewage outlet, the pollution level of the river after mixing is equal to the sum of pollution level of sewage and river before mixing.

Nevertheless, the first model ignored the degradation of pollutants along the river runoff. There comes the second, called the \emph{one-dimensional water quality model}, is designed for describing this phenomenon.
However, the one-dimensional water quality model omits the diffusion of pollutants at the river cross-section and considers no sewage inflows.
In this case, mark an arbitrary location along the river as $i_1$, and denote the pollutant concentration value of river cross-section at location $i_1$ as $C_{i_1}$. Then the pollutant concentration value at location $i_2$, which is downstream of location $i_1$, can be represented as $C_{i_2}=C_{i_1}\exp^{-\kappa D(i_1,i_2)}$, where $D(i_1,i_2)$ is the distance from location $i_1$ to $i_2$, and constant $\kappa$ represents the pollutants degradation rate per unit distance along the river.
Therefore, term $\exp^{-\kappa D(i_1,i_2)}$ could be viewed as the residual rate of pollutants between locations $i_1$ and $i_2$.
Accordingly, for any location $i_3$ at the downstream of location $i_2$, the pollutant concentration value at here is $C_{i_3}=C_{i_1}\prod_{i=i_1}^{i_3-1} \exp^{-\kappa D(i,i+1)} = C_{i_1}\exp^{-\kappa D(i_1,i_3)}$.

We will show in Section \ref{sec3} how these two models are jointly applied to our sewage discharge problems to describe the water quality at a certain location along the linear river.

\subsection{Games with Transferable Utility and Externalities}\label{sec2.2}

In the context of cooperative game theory, a \emph{cooperative game with transferable utility} (TU-game) describes a game situation where players can generate and transfer certain payoffs by cooperation. 
We use a pair of notations $(N,v)$ to represent a TU-game, where $N:=\{1,\dots,n\}$, referred to as the \emph{player set}, is a nonempty while finite set with $N$ being the grand coalition and each subset $S$ of $N$ being a sub-coalition, and $v:2^N \rightarrow \mathbb{R}$, referred to as the \emph{characteristic function}, maps each coalition $S\subseteq N$ to a value of $v(S)$ that represents the maximum total profit for players in $S$ to accomplish certain work through cooperation.

To seek possible cooperation among all players in $N$ so to achieve global optimum, a TU-game needs to announce in advance an acceptable $n$-dimensional \emph{allocation vector} $(\alpha_i)_{i\in N}\in \mathbb{R}^n$ with $\alpha_i$ being the profit allocated to each player $i\in N$.
Among various means of defining acceptable allocation vectors in the literature, the concept of \emph{core} proposed by \cite{shapley1969market} is one of the most important allocation concepts under which all players in $N$ are willing to cooperate.
To be specific, the core of game $(N,v)$ is defined as the set of allocation vectors $(\alpha_i)_{i\in N}\in \mathbb{R}^n$ satisfying an budget balanced constraint, i.e. $\sum_{i\in N}\alpha_i=v(N)$, so that all the profit generated by grand cooperation is totally shared by the players, as well as a set of coalition stability constraints is satified, i.e., $\sum_{i\in S}\alpha_i\geq v(S), ~\forall S \subset N$, so that no sub-coalition has incentive to deviate to seek higher profit.
To summarize, we can mathematically define the core as follows:
\begin{equation}\label{core}
    \text{Core}(N,v)=\left\{(\alpha_i)_{i\in N}\in \mathbb{R}^n:\sum_{i\in N}\alpha_i=v(N),~\sum_{i\in S}\alpha_i\geq v(S),~\forall S\subset N\right\}.
\end{equation}

As will be explicitly analyzed in Sections \ref{sec3}, \ref{sec4} and \ref{sec5}, the coalition value $v(S)$ for some $S \subset N$ in certain cooperative game (e.g., the the sewage discharge game studied in this paper) would be severely affected by the decisions made by external game players $N \setminus S$.
To this end, besides TU-game we also need to introduce the concept of \emph{game with externalities} for preliminaries, where, as interpreted by \cite{neumann1944theory}, the coalition value of $S$ should be defined as the maximum total profit that $S$ can guarantee its members.

Accordingly, \cite{chander1997the} propose the $\gamma$-characteristic function $v^\gamma(\cdot)$, that represents the maximum total profit $S$ could gain when all players in $N \setminus S$ are individually making myopic optimization decisions.
The value of $v^\gamma(S)$ for each $S\subseteq N$ could be formally defined as
$$
\begin{aligned}
    v^\gamma(S) = & \max \sum_{i\in S}\mathcal{U}_i ((e_i)_{i\in S}\cup (e_i^*)_{i\in N\setminus S}) & & \\
    s.t. \quad & e_j^*=\mathop{\arg\max}_{e_j\in E_j}\left\{ \mathcal{U}_j \left((e_i^*)_{i\in N\setminus (S\cup j)}\cup (e_i)_{i\in S}\cup e_j\right)\right\}, & & \text{for all}\ j\in N\setminus S, \\
    & e_i\in E_i, & & \text{for all}\ i\in S,
\end{aligned}
$$
where, for a player $i\in N$, $e_i$ represents his decision choice, $E_i$ represents his decision profile, $\mathcal{U}_i\big((e_i)_{i\in N}\big)$ denotes his utility when the decisions of players in $N$ are $(e_i)_{i\in N}$, and $e_j^*$ is the myopic optimal decision for outsider $j\in N\setminus S$ given that players in $N\setminus \{S \cup j\}$ are also making myopic decisions and players in coalition $S$ are making group optimal decisions.

\subsection{Notations Specified for Optimization and Cooperation in a Line}
Our main focus in this paper is to analyze the optimization and cooperation problems in sewage discharging in a line along river. Below we simply refer to a player as \emph{a node} in player (node) set $N$ unless otherwise specified.
For ease of exposition, below in Table \ref{tab1} we summarize some notations and expressions that will be used throughout the remainder of this paper.
\begin{table}[ht!]
    \renewcommand\arraystretch{1.5}
    \caption{Notations and Expressions}
    \centering
    \begin{tabular}{cp{14cm}}
    \hline
        Notations & \makecell[c]{Meanings}  \\ \hline
        $N$ & the grand coalition $N=\{1,\dots,n\}$, where nodes are successively numbered from upstream to downstream;\\
        $n,~s,~t$ & the respective number of players in coalitions $N,~S,~T$;\\
        $S\cup i$ & the union of a coalition $S$ and a singleton coalition $\{i\}$;\\
        $S\setminus i$ & A coalition $S$ minus a singleton coalition $\{i\}$;\\
        $S^h$ & the $h^{\text{th}}$ upstream node in coalition $S$;\\
        $\min S$ & also denoted as $S^1$, the $1^{\text{st}}$ upstream (the head) node in $S$, i.e., $S^1=\min\{i:\forall i\in S\}$;\\
        $\max S$ & also denoted as $S^s$, the $s^{\text{th}}$ upstream (the tail) node in $S$, i.e., $S^s=\max\{i: \forall i\in S\}$;\\
        $\prec(\succ)$ & is upstream (downstream) of, for instance, $i\prec T$ implies that $i<j$ for all $j\in T$, i.e., $i$ is upstream; $S\prec T$ implies that $i_1<i_2$ for all $i_1\in S$ and $i_2\in T$;\\
        $\big[i_1:i_2\big]$ & a consecutive segment of nodes of $N$ starting from node $i_1$ and ending at node $i_2$, i.e., $\big[i_1:i_2\big] = \{i_1,i_1+1,\dots,i_2\}$;\\
        $S^{[1:s]}$ & a consecutive segment of nodes of $N$ starting from node $S^1$ and ending at node $S^s$, i.e., $S^{[1:s]} = [S^1:S^s]$, which may include nodes not in $S$;\\
        $S^{< i}$ &a subset of $S$ containing all upstream nodes of $i$, i.e., $S^{< i} = \{j\in S: j <  i,~\forall i\in N\}$;\\
        $S^{> i}$ &a subset of $S$ containing all downstream nodes of $i$, i.e., $S^{> i} = \{j\in S: j >  i,~\forall i\in N\}$.\\  \hline
    \end{tabular}
    \label{tab1}
\end{table}

\section{Sewage Discharge Problem}\label{sec3}
In order to study the cooperative sewage discharge game along a river, it is essential to first analyze the optimization problem of sewage discharge (SDP) for each coalition of firms.
In Section \ref{sec3.1}, we will model and analyze the global SDP $v(N)$ for the universal set of firms (i.e., the grand coalition $N$), and accordingly in Section \ref{sec3.2}, we will propose an algorithm based on greedy strategy that optimally solves the global SDP in polynomial time.
Building upon the results derived from the SDP for the grand coalition, in Section \ref{sec3.3} we will further analyze the SDP $v(S)$ for a sub-coalition $S \subset N$ and propose a polynomial time solvable algorithm for it as well.

\subsection{Model for Grand Coalition}\label{sec3.1}

Consider the case where $n$ nodes (e.g., firms) are sequentially located along a linear river from upstream to downstream. Under grand coalition, there is a central authority who can decide the sewage discharge quantity $x_i$ for each node $i\in N$. The sewage discharge is related with profit generating activities such as production.
We assume that the profit generated by node $i$ is a monotonically increasing, differentiable and concave function of $x_i$, which is given by $f_i(x_i)$.
Without loss of generalities, we also assume that no sewage discharge makes zero profit, i.e., $f_i(0)=0$.

In terms of sewage discharging, there are two well-recognized types of restrictions, namely, \emph{the sewage discharging bounds} and \emph{the river pollution tolerances}.
To be specific, first, it is necessary to set an interval of sewage discharge level $\big[a_i,u_i\big]$ for each node $i \in N$, where $a_i$ can be interpreted as a basic sewage discharge level for maintaining its minimum daily production level, while $u_i$ can be interpreted as a maximum sewage discharge level related with its production capability upper bound.
Second, due to practical reasons such as governmental regulations for river pollution tolerances, the pollution level at each sewage outlet $i \in N$ has to be constrained within a threshold $b_i$.
To investigate different feasible sewage discharge schemes, we have to make a mild assumption on the two sewage discharging restrictions, that is, the basic sewage discharge level $a$ is low enough compared with the river pollution tolerance $b$. For instance, the minimum sewage discharge level required by a firm is $5$; however if the river pollution tolerance at the location of this firm was limited less than $5$, then it has to go out of business.

Let $p'_i$ and $p_i$ denote the pre-discharging and post-discharging pollution levels at node $i$ before and after sewage discharging, respectively. In this paper, we consider a simple hydrologic regime. Based on the sewage mixing and degradation processes introduced in the preliminaries of water quality models in Section 3.1, we have that, for each $i \in [2:n]$,
$$
p'_{i} = k_{i-1} p_{i-1} ~(0< k_i \leq 1), \text{  and   } p_i = p'_i+x_i \leq b_i,
$$
where $k_{i-1} = \exp^{-\kappa D(i-1,i)}$ is the residual rate of pollutants between nodes $i-1$ and $i$, so pollution level at node $i$ is the mixture of residual pollution from node $i-1$ and the discharge level from node $i$ (see more descriptions in Section \ref{sec2.1}).
While for node $1$, without loss of generalities, we simply let $k_0=1$, implying that the distance between nodes ``$0$" and $1$ is zero, and denote the initial pollution level, as well as the post-discharging pollution level, at node $0$ as $b_0$. We have that $p'_{1} = k_{0} b_{0}$ and $p_1 = p'_1+x_1 = k_0b_0 + x_1 \leq b_1$.

For ease of expression, we use a five-member tuple $\mathcal{T}_N:=(b_i,k_i,a_i,u_i,f_i)_{i\in N}$ to define the main parameters of the SDP for $n$ nodes along a river, and describe the SDP for the grand coalition $N$ as SDP$(b_0,N)$.
In the myopic situation where players are myopically maximizing their own profits, it is intuitive that the resulting \emph{myopic post-discharging pollution level}, denoted as $d_i$ for each $i \in N$, can be written in a close form as
$$
d_i=\min\{u_i+k_{i-1}d_{i-1}, b_i\},~ d_0=b_0,
$$
implying that each player is discharging up to its sewage discharging upper bound $u_i$ or is bounded by the river pollution tolerance $b_i$.
Accordingly, we can derive the \emph{myopic sewage discharge solution} $\Theta=(\theta_i)_{i\in N}$ for nodes along the river as $\theta_i = d_i - k_{i-1}d_{i-1}$, for each $i\in N$.

However, as a grand coalition $N$, the objective is to globally maximize the total profit $\sum_{i\in N}f_i(x_i)$ by deciding sewage discharge level jointly.
In this case, we can formulate the SDP$(b_0,N)$ as follows,
\begin{align}\label{model1}
    v(N)=\max \quad & \sum_{i\in N}f_i(x_i) & \\
    s.t. \quad & \sum_{j=1}^i K_j^i x_j + K_1^i b_0\leq b_i,  & \forall i\in N, \notag  \\
    & a_i\leq x_i \leq u_i, & \forall i\in N, \notag
\end{align}
where $K_{i_1}^{i_2}:=\prod_{i=i_1}^{i_2-1} k_i$, for $1 \leq i_1 < i_2 \leq n$, represents the residual rate of pollutants from upstream node $i_1$ to downstream node $i_2$. $K_i^i=1$, for each $i \in N$, is associated with the sewage discharging quantity $x_i$ at node $i$. Note that given the values of $k$'s, the $K$'s are fixed parameters.
As a result, for each $i \in N$, the first term $\sum_{j=1}^i K_j^i x_j$ in constraint $\sum_{j=1}^i K_j^i x_j + K_1^i b_0\leq b_i$, in fact denotes the cumulative pollution level at node $i$ due to sewage discharged by nodes in $[1:i]$, while the second term $K_1^i b_0$ is simply the residual pollution level at node $i$ caused by the initial pollution level $b_0$. 
Due to the pollution tolerance constraint, the sum of the two terms should be no larger than the river pollution tolerance $b_i$.
In addition, constraint $a_i\leq x_i \leq u_i$ regulates the sewage discharging bounds for nonlinear objective of each $x_i$.

Noticing that in SDP (\ref{model1}) all constraints are linear, however the optimization is still complicated due to the non-linearity of profit function $f_i(\cdot)$ for each $i \in N$. Recall that the $f_i(\cdot)$ function is monotonically increasing and concave. Fortunately, we can derive several optimality conditions for the SDP of the grand coalition as presented in Theorem \ref{thm1} and Lemma \ref{lem1}, which will help us to design the optimal algorithm. 
To ease notation, we let the term $c_i(X)=\sum_{j=1}^i K_j^i x_j + K_1^i b_0, \text{ for each $i \in N$}$, indicate the post-discharging pollution level at node $i$ under some given sewage discharge quantity $X=(x_i)_{i\in N}$, and denote $f'_i(x_i)$ as the first order derivative of $f_i(x)$ at $x=x_i$.

\begin{theorem}\label{thm1}
    Given an optimal solution $X^*=(x_i^*)_{i\in N}$ for the SDP$(b_0,N)$, we have that the optimal post-discharging pollution level $c(X^*)$ satisfies
    $$c_i(X^*)\leq d_i, \text{ for each }i \in [1:n-1],\text{~and~} c_n(X^*)=d_n.$$
\end{theorem}

Recall that $d_i$ is the post-discharging pollution level of node $i$ under the myopic setting. As we can see from Theorem \ref{thm1}, in the optimal scheme, the post-discharging pollution level $c_i(X^*)$ at node $i \in \big[1:n-1\big]$ is bounded by $d_i$, while $c_n(X^*)$ at node $n$ is equal to $d_n$. In other words, node $n$ would always discharge up to its sewage discharging upper bound $u_n$ or reaches the river pollution tolerance $b_n$. As a result, Theorem \ref{thm1} shows that global optimization can achieve less pollution and more benefits for the grand coalition.

In addition to the optimality conditions for the post-discharging pollution levels $c(X^*)$, in Lemma \ref{lem1}, we further describe the optimality conditions for the sewage discharging quantity $X^*$ by comparing the marginal benefits of discharging an additional unit of sewage within two adjacent nodes.

\begin{lemma}\label{lem1}
    Given an optimal solution $X^*=(x_i^*)_{i\in N}$ for the SDP$(b_0,N)$, for any two adjacent nodes $i$ and $i+1$ in $\big[1:n\big]$, we have that
    \begin{itemize}
        \item if inequality $f'_i(x_i^*)<k_i \cdot f'_{i+1}(x_{i+1}^*)$ holds, then at least one of the following two conditions \big\{$x_i^*=a_i$, $x_{i+1}^*=u_{i+1}$\big\} is true;
        \item if inequality $f'_i(x_i^*)>k_i \cdot f'_{i+1}(x_{i+1}^*)$ holds, then at least one of the following three conditions \big\{$c_i(X^*)=d_i$, $x_i^*=u_i$, $x_{i+1}^*=a_{i+1}$\big\} is true.
    \end{itemize}
\end{lemma}

To be more specific, Lemma \ref{lem1} indicates that for two adjacent nodes $i$ and $i+1$, by taking $k_i$, the residual rate of pollutants, from node $i$ to node $i+1$ into consideration, the central authority should assign as much discharging quantity as possible to the node with higher normalized marginal benefit, i.e., with higher value of $f'_i(x_i^*)$ or $k_i \cdot f'_{i+1}(x_{i+1}^*)$.
When the upstream node $i$ has lower normalized marginal benefit $f'_i(x_i^*)$, its optimal discharging quantity $x_i^*$ is set to be its lower bound $a_i$ unless the discharging quantity of the downstream node $i+1$ has already reached its upper bound $u_{i+1}$. 
When the upstream node $i$ has higher normalized marginal benefit, its optimal discharging quantity $x_i^*$ is set to be its upper bound $u_i$, or it should act in a myopic way such that its post-discharging pollution level 
reaches $d_i$, unless the discharging quantity of the downstream node $i+1$ already reaches its lower bound $a_{i+1}$.

Motivated by Lemma \ref{lem1}, we denote term $K_1^i f'_i(\cdot)$ as the \emph{adaptive marginal benefit} (AMB) for node $i \in N$. The central authority should assign more discharging quantities to a node with higher AMB. It is clear that this AMB plays an important role to maximizing the total profit for the grand coalition. This inspires the greedy-like strategy in the next section.

\subsection{Algorithm with Greedy Strategy}\label{sec3.2}
In this section, by utilizing the properties of AMB we develop a greedy-like algorithm to solve the SDP for the grand coalition.
The key idea is to first construct a basic feasible sewage discharge scheme, and then greedily increase sewage discharging quantities for nodes with highest AMB. Indeed the optimal discharge level of each node can be no less than it is at any stage of our algorithm.

We aforementioned in Theorem \ref{thm1} that the myopic pollution level $d_i$ is the upper bound of pollution level at any node $i\in N$.
For the SDP, applying this in the greedy strategy, we can obtain that if the pollution level at a node $i\in N$ reaches $d_i$, then discharging quantities of all upstream nodes of $i$ cannot be increased anymore.
Thus, we raise the following lemma and definition.

\begin{lemma}\label{lem2}
    Given an optimal solution $X^*=(x_i^*)_{i\in N}$ for the SDP$(b_0,N)$, if there exists some node $j\in N$ such that $c_j(X^*)=d_j$, then $(x_i^*)_{i\in [1:j]}$ is also an optimal solution of SDP$\big(b_0,[1:j]\big)$.
\end{lemma}

\begin{definition}\label{def1}
    The grand coalition $N$ is decomposable if there is a node $j\in [1:n-1]$ with pollution level $c_j(X^*)=d_j$ under an optimal solution $X^*$ of the SDP$(b_0,N)$.
\end{definition}

Intuitively, Lemma \ref{lem2} indicates that, under the greedy strategy of solving SDP (\ref{model1}), if the pollution level at a node $j\in N$ is increased to $d_j$, then the greedy strategy solution can be decomposed into two parts: one is the optimal solution of SDP$\big(b_0,[1:j]\big)$ and another is the initial constructive solution of a sub-optimization problem SDP$\big(d_j,[j+1:n]\big)$. 
On this basis, the first segment $[1:j]$ has reached optimality. For all nodes in $[j+1:n]$, their discharging quantities can continue to increase following the greedy strategy. We can repeat the procedure to further decompose $[j+1:n]$, until the pollution level at $n$ reaches $d_n$. Then the optimal discharge solution of SDP (\ref{model1}) is obtained.
In brief, if the grand coalition $N$ is decomposable, which means that there exists $j\in [1:n-1]$, the grand-optimization problem SDP$(b_0,N)$ can be decomposed into two sub-optimization problems: SDP$\big(b_0,[1:j]\big)$ and SDP$\big(d_j,[j+1:n]\big)$.

For illustration, we first consider a special case where, for each node $i \in N$, the profit function $f_i(x_i)$ is linear in $x_i$, i.e., the AMB of node $i$---$K_1^i f'_i(a_i), \forall a_i\leq x_i\leq u_i$---is constant. 
Below we summarize the framework of a greedy algorithm solving SDP (\ref{model1}) in this linear case.

\begin{itemize}
    \item Firstly, for initialization, construct a basic while feasible sewage discharge scheme $(x_i)_{i\in N}$ for SDP (\ref{model1}).
    According to Lemma \ref{lem1}, we can see that it is possible for some node $i$ to discharge only $a_i$ in the optimal case.
    Thus, for ease to access, we simply let the basic sewage discharge scheme start from $X$ where $x_i=a_i$ for each $i \in N$.
    Let $I^*$ be an empty set and later on iteratively add nodes with the highest AMBs.
    \item Secondly, iteratively identify a node $i^*=\arg \max_i\{K_1^i f'_i(a_i): \forall i\in N \setminus I^*\}$ with the highest AMB and let $I^*= I^* \cup i^*$.
    Due to the constant margin $f'_i(x_i)$, let node $i^*$ discharge as much additional sewage as possible, such that its maximum sewage discharge level $u_{i^*}$ is not exceeded and the resulting post-discharging pollution level of each node $j\geq i^*$ does not exceed the corresponding myopic pollution level $d_j$.
    Thus, the discharging quantity of node $i^*$ is at most increased by
    $$\delta^*=\max\{\delta: x_{i^*} + \delta \leq u_{i^*}, \text{ and } c_j(X) + K_{i^*}^j \delta \leq d_j,~\forall j \geq i^* \}.$$
    The resulting discharging quantity of node $i^*$ is $x_i + \delta^*$, and it could be proved optimal under certain situations of global optimization.
    Accordingly, we update the sewage discharge scheme $X$ by replacing $x_i$ with $x_i + \delta^*$. Then we go to the next iteration.
    \item Thirdly, the algorithm would stop whenever the post-discharging pollution level at node $n$ reaches $d_n$, since no more sewage discharge is allowed for any $i \in N$.
    As we can see from the second step, the algorithm stops within at most $n$ iterations.
\end{itemize}

Nevertheless, for the general case where $f_i(\cdot)$ is concave and maybe nonlinear for each node $i \in N$, the above framework is not applicable. This is because AMB, which is equal to $K_1^i f'_i(x_i)$, keeps updating as the sewage discharge $x_i$ changes.

To utilize the greedy strategy shown above as in the linear case, we still let the basic sewage discharge scheme starts from $X = (a_i)_{i\in N}$, and let $I^* = \varnothing$ initially.
Instead of solely increasing the discharging quantity of a single node outside $I^*$ having the highest AMB in each iteration, our greedy strategy increases the discharging quantities of a set of nodes having the same while highest AMB, such that their AMBs are maintained at the same level, and we gradually expand the set over the iterations.

For illustration, we start with the value $f'_i(a_i)$ at the beginning, denote the node with the $m^{\text{th}}$ highest AMB as $i_m^*$. At the first iteration we can let $I^* = \{i_1^*\}$, and then let node $i_1^*$ discharge as much additional sewage as possible such that, subject to discharging restrictions, its AMB declines to $K_1^{i_2^*}f'_{i_2^*}(a_{i_2^*})$, i.e., the AMB of node $i_2^*$.
Then let $I^*=\{i_1^*,i_2^*\}$ and we will discharge $I^*$ simultaneously so their AMB are maintained at the same level. Note that the discharge quantities of $i_1^*$ and $i_2^*$ are not necessarily same.
In the $m^{\text{th}}$ iteration, we have $I^* = \big\{i_1^*, i_2^*, \cdots, i_m^*\big\}$, and then let nodes in $I^*$ discharge simultaneously, complying with discharging restrictions, until their AMBs decline to $K_1^{i_{m+1}^*}f'_{i_{m+1}^*}(a_{i_{m+1}^*})$, i.e., the initial AMB of node $i_{m+1}^*$.
At each iteration, if there is a node $i\in I^*$ whose discharging quantity reaches upper level $u_i$, then stop all nodes in $I^*$ from discharging, and remove $i$ from $I^*$ in the next iteration. 

As we can see from the above descriptions, the greedy strategy is increasing the sewage discharging quantity \emph{layer by layer} until the SDP is optimally solved.
During the $m^{\text{th}}$ iteration, the nodes dischargeable at layer $m$ are given by $I^*$. The synchronization of layer-increment can be described as the \emph{balanced equation}, where we claim that, when the $m^{\text{th}}$ iteration terminates, the discharging quantity $(x_i)_{i\in I^*}$ will satisfy
$$
    K_1^{j_1}f'_{j_1}(x_{j_1})=K_1^{j_2}f'_{j_2}(x_{j_2}), \text{ for any two nodes $j_1,~j_2\in I^*$,}
$$
and the terminating condition of this iteration is decided by the initial AMB of node $i_{m+1}^*$.
Therefore, the incremental quantity for each node in $I^*$ can be calculated exactly.

As a result, our greedy strategy would directly solve the exact optimal incremental discharging quantity in each iteration, instead of gradually increase discharging quantity until not being able to increase anymore. 
For ease of exposition, let $g:= (f')^{-1}$ be the inverse function of $f'$. We formalized the greedy-based Layer-by-Layer-Incremental algorithm (simply as LLI algorithm) as follows.

~\
\begin{breakablealgorithm}
	\renewcommand{\algorithmicrequire}{\textbf{Input:}}
	\renewcommand{\algorithmicensure}{\textbf{Output:}}
    \caption{\textsc{LLI Algorithm}}
    \label{alg1}

	\begin{algorithmic}[1]
		\Require $(b,k,a,u,f)_i$ for all nodes in $N$, initial pollution level $b_0$, and myopic pollution level $(d_i)_{i\in N}$
		\Ensure Discharge solution $X$
        \State Start from set $\Gamma=\varnothing$ and initial discharge $X=(a_i)_{i\in N}$
        \Repeat
        \State Update $\big(c_i(X)\big)_{i\in N}$ by $c_i(X)=\sum_{j=1}^iK_j^i x_j+K_1^i b_0$
        \State Obtain $I^*:=\arg\max\{K_1^i f'_i(x_i),~ \forall i\in N\setminus \Gamma\}$ and let $i^*:=\min\{i, ~ \forall i\in I^*\}$
        \State Define $(\Delta x_i)_{i\in I^*}$ such that $\Delta x_j= g_j\big(f'_{i^*}(x_{i^*}+\Delta x_{i^*})/K_{i^*}^{j}\big)-g_j\big(f'_{i^*}(x_{i^*})/K_{i^*}^{j}\big)$ for all $j\in I^*$
        \If{$I^*\subset N\setminus \Gamma$} 
            \State Let $\xi:=\max\{K_1^i f'_i(x_i), ~ \forall i\in N\setminus (\Gamma \cup I^*) \}$
            \State Calculate $\sigma_1:=g_{i^*}(\xi/K_1^{i^*})-x_{i^*}$
        \Else
            \State Let $\sigma_1$ be a large enough number, i.e., larger than $\max\{u_i-a_i,~ i\in N\}$
        \EndIf
        \State Calculate $\sigma_2:=g_{i^*}\big(f'_{j^*}(u_{j^*})/K_{i^*}^{j^*}\big)-x_{i^*}$, where $j^*=\arg\max\{K_1^j f'_j(u_j), ~ \forall j\in I^*\}$
        \State Calculate $\sigma_3:=\max\left\{\Delta x_{i^*}:\sum_{i\in I^*}^{i\leq j}K_i^j \Delta x_i \leq b_j-c_j(X), ~ \forall j\geq i^*\right\}$
        \State Get $\Delta x_{i^*}:=\min\{\sigma_1,\sigma_2,\sigma_3\}$, and let $x_j:=x_j+\Delta x_j$ for all $j\in I^*$

        \If{$\Delta x_{i^*}=\sigma_2$}
            \State Let $\Gamma:=\Gamma \cup j^*$
        \ElsIf{$\Delta x_{i^*}=\sigma_3$}
            \State Obtain $l^*$ in $[i^*:n]$ with $\sum_{i\in I^*}^{i\leq l^*}K_i^{l^*}\Delta x_i= b_{l^*}-c_{l^*}(X)$
            \State Let $\Gamma:=\Gamma \cup [1:l^*]$
        \EndIf
        
        \Until{$\Gamma$ equals to $N$}
		\State \textbf{return} $X$
	\end{algorithmic}  
\end{breakablealgorithm}

~\

Below in Proposition \ref{prop2} we formally prove that the LLI algorithm can optimally solve the SDP for the grand coalition $N$.
\begin{proposition}\label{prop2}
    The final sewage discharge scheme $X$ output by the LLI algorithm is optimal to the SDP for the grand coalition $N$, formulated by the nonlinear programming in (\ref{model1}).
\end{proposition}

In the LLI algorithm, we can find that the time complexity of solving for $\sigma_1$, $\sigma_2$, $\sigma_3$ depend on the form of profit functions, e.g., power function, logarithmic function etc.
For instance, in the special case where $f_i(x_i)$ is linear in $x_i$ for all $i\in N$, the time complexity of deriving $\sigma_3$ in each iteration would simply be $\mathcal{O}(n)$ by solving $\min \left\{\big(b_j-c_j(X)\big)/ \sum_{i\in I^*}^{i\leq j}K_i^{j}, ~\forall j\geq i^*  \right\}$.

Below in Lemma \ref{lem3} we further show that in the general case, where $f'_i(x_i)$ has analytical solution and $f_i(x_i)$ is nonlinear in $x_i$ for all $i\in N$, the LLI algorithm has polynomial time complexity.

\begin{lemma}\label{lem3}
    The number of iterations in step $2\sim 21$ is less than $\mathcal{O}(n)$. The LLI algorithm has the time complexity at most $\mathcal{O}(n^3)$.
\end{lemma}

In the end, we want to remark that the LLI algorithm is also applicable to the case where $f'(x)$ does not have a closed-form, while at the cost of sacrificing precision or computational time of solving for $\sigma_1$, $\sigma_2$, $\sigma_3$.

\subsection{Model and Algorithm for Sub-coalition}\label{sec3.3}

When only a subset of players, $S \subset N$, choose to cooperate, the resulting optimization problem is an SDP for sub-coalition $S$.
In this case, compared with the SDP for the grand coalition $N$, the objective is now changed to maximizing $\sum_{i\in S}f_i(x_i)$, while the constraints are not simply the sewage discharging bounds and river pollution tolerances, as shown in programming (\ref{model1}), for all nodes in $S$. This is because the decisions of nodes in $N \setminus S$ would also have external impacts on the decisions of $S$, so we must take their decisions in the consideration. As common practice, upon generating coalition $S$, each node in $N \setminus S$ makes discharge decision according to the concept of $\gamma$-characteristic function (see Section \ref{sec2.2}). Therefore each node outside $S$ is selfish and wants to discharge as much sewage as possible so that its own profit is maximized.

Recall that $S^1$ and $S^s$ respectively represent the head and the tail nodes in $S$, and $S^{[1:s]}$ is a consecutive set of nodes along a river starting from $S^1$ and ending at $S^s$ which may include node not in $S$.
To formulate the SDP for $S$, we have the following observations.
First, each node $i$ in $\big[1:S^1-1\big]$ and $\big[S^s+1:n\big]$, i.e., all upstream and downstream nodes to $S$, would act in a myopic way by exactly discharging sewage to the optimal myopic discharging quantity $\theta_i$.
Second, some node $i$ in $S$ may have incentive to seek cooperation with its downstream nodes in $S$ by saving its discharging quantity.
Third, each node $i$ in $S^{[1:s]} \setminus S$ would selfishly discharge to its upper bound $u_i$ or discharge up to its river pollution tolerance $b_i$. Note that in fact it has the chance to discharge more than the optimal myopic discharging quantity $\theta_i$ since its upstream nodes in $S$ might have the incentive to discharge less to seek cooperation with its downstream nodes in $S$.

Based on the above observations, we can mathematically formulate the SDP for a sub-coalition $S \subseteq N$ as the following nonlinear programming
\begin{align}\label{model2}
    v(S)=\max \quad & \sum_{i\in S}f_i(x_i)  & &\\
    s.t.\quad & \sum_{j=S^1}^i K_j^{i}x_j+ K_{S^1-1}^{i}d_{S^1-1} \leq b_i, & & \forall i\in S^{[1:s]}, \notag \\
    & a_i\leq x_i \leq u_i,& & \forall i\in S, \notag \\
    & x_i=\min\{u_i,~b_i-k_{i-1}c_{i-1}(X)\}, & & \forall i\in S^{[1:s]}\setminus S. \notag
\end{align}

As we can see from (\ref{model2}), all nodes in $S^{[1:s]}$ are constrained by the river pollution tolerances such that, for each $i$, the cumulative residual pollution level from all upstream nodes including node $i$, is bounded by $b_i$.
The feasible region of discharging quantity $x_i$, for each $i \in S$, is simply the interval $\big[a_i,u_i\big]$, while the discharging quantity $x_i$, for node $i\in S^{[1:s]}\setminus S$, is $\min\{u_i,~b_i-k_{i-1}c_{i-1}(X)\}$. In other words, it myopically maximize its profit with no cooperation.
Similar to SDP$(b_0,N)$, for ease of notation, we describe the SDP for sub-coalition $S$ as SSDP$(d_{S^1-1},S,N)$ since the residual pollution of node $S^1-1$ must be at the myopic level $d_{S^1-1}$.

In a special case where $S$ is a consecutive subset of $N$, we can see that the resulting SSDP$(d_{S^1-1},S,N)$ can be solved by the LLI algorithm for SDP$(d_{S^1-1},S)$. However, in a general case where $S$ is inconsecutive, as the supports of $(x_i)_{i\in S^{[1:s]}\setminus S}$ in SSDP$(d_{S^1-1},S,N)$ are computed by nonlinear functions specified in the last constraint, the LLI algorithm is no longer applicable.
To this end, below in Lemma \ref{lem4} we characterize the optimal value of $(x_i)_{i\in S^{[1:s]}\setminus S}$ in the SSDP$(d_{S^1-1},S,N)$. 
This structure combining the LLI algorithm allows us to provide a dynamic programming to recursively solve the SSDP$(d_{S^1-1},S,N)$ in Algorithm \ref{alg2}.

\begin{lemma}\label{lem4}
    Given an optimal solution $(x_i^*)_{i\in S^{[1:s]}}$ for the SSDP (\ref{model2}), the discharging quantity $x_i^*$ for each $i\in S^{[1:s]}\setminus S$ is equal to either $\theta_i$ or $u_i$.
\end{lemma}

With Lemma \ref{lem4}, SSDP (\ref{model2}) can be formulated as a mixed-integer programming, that enumerates possible discharging quantities of all nodes in $S^{[1:s]}\setminus S$. Moreover, Lemma \ref{lem4} has two implications: First, under some cases, $(u_i)_{i\in S^{[1:s]}\setminus S}$ is not a feasible discharge solution. Second, not all feasible discharging quantities are plausible, e.g., it is not rational that a node $i\in S^{[1:s]}\setminus S$ discharges $\theta_i$ with $\theta_i<u_i$ while the pre-discharging pollution level at $i$ is less than $k_{i-1}d_{i-1}$.

Under the first implication, we define $v'(S)$ as follows:
$$
\begin{aligned}
    & v'(S)= \sum_{i\in S} f_i(x_i^*) \\
    & \begin{aligned}
        (x_i^*)_{i\in S^{[1:s]}}= \quad & \arg\max ~ \sum_{i\in S}f_i(x_i) + \sum_{i\in S^{[1:s]}\setminus S}\beta\cdot x_i  & &  \\
        s.t.\quad & \sum_{j=S^1}^i K_j^{i}x_j+ K_{S^1-1}^{i}d_{S^1-1} \leq b_i, & & \forall i\in S^{[1:s]},  \\
        & a_i\leq x_i \leq u_i,& & \forall i\in S^{[1:s]},  \\
    \end{aligned}
\end{aligned}
$$
where $\beta$ can be any large number such that $\beta> \max \{K_1^if'_i(a_i),i\in S\}$, i.e., the largest initial AMB among all nodes.
This programming can be regarded as that the central authority give priority to the nodes in $S^{[1:s]}\setminus S$ so they can discharge as much as possible. As a result, this formulation can be regraded as an SDP for a consecutive set $S^{[1:s]}$. 
Therefore, it is obvious that the optimal solution of $v'(S)$ can be obtained by the LLI algorithm. Moreover, we have $v(S)\geq v'(S)$ with equality when the discharging quantity of each node $i\in S^{[1:s]}\setminus S$ is equal to $u_i$ under both problems.

Under the second implication, if there exists some node $i\in S^{[1:s]}\setminus S$ discharges less than $u_i$, then it must be the case that the post-discharging pollution level at $i-1$ reaches $d_{i-1}$. Analogously inspired by Definition \ref{def1}, we can decompose SSDP$(d_{S^1-1},S,N)$ into different combinations of subproblems, and find the best decomposition method to obtain the optimal discharge solution.
A method of decomposing SDP is given in Lemma \ref{lem5}, that is, decomposing the SSDP$(d_{S^1-1},S,N)$ into two sub-SDPs for coalitions $S_1$ and $S\setminus S_1$, where $S_1\subset S$.

\begin{lemma}\label{lem5}
    Given a coalition $S$, for each subset $S_m\subset S$, we have that $v(S)\geq v(S_m)+v(S\setminus S_m)$.
\end{lemma}

Combine the programming of $v'(S)$ and Lemma \ref{lem5}, $v(S)$ can be recursively solved by the dynamic programming as:
$$
v(S)=\max\left\{v'(S),\max_{S_m\subset S,S_m\not= \varnothing}\big\{v(S_m)+v(S\setminus S_m)\big\}\right\}.
$$
However it is computationally intractable to enumerate all subsets of $S$ and choose the best combination.
Nevertheless, we ask the question ``do we need to try all possible decompositions of $S$?" Fortunately, the answer is no. It turns out that we only need to care about consecutive partitions. 
More specifically, we partition $S$ into consecutive and disjoint subsets $S_1,\dots,S_l$ with each subset only containing consecutive nodes and the size of each subset as large as possible, e.g. $\{1,2,4,5\}$ is partitioned into $\{1,2\}$ and $\{4,5\}$. Then we only need to consider decomposing SSDP$(d_{S^1-1},S,N)$ into SSDP$\left(d_{S^1-1},\bigcup_{m=1}^{m_1} S_m,N\right)$ and SSDP$\left(d_{S_{m_1}^1-1},\bigcup_{m=m_1+1}^{l} S_m,N\right)$ for all $m_1\in [1:l-1]$. In each decomposition pair $\left(\bigcup_{m=1}^{m_1} S_m,\bigcup_{m=m_1+1}^{l} S_m\right)$, it is obvious that any node $i\in \big[(\max S_{m_1})+1:(\min S_{m_1+1})-1\big]$ will myopically discharge sewage by quantity $\theta_i$.

\begin{theorem}\label{thm2}
    Given a coalition $S$, the optimal objective value of SSDP$(d_{S^1-1},S,N)$ can be obtained by recursion
    $$
    v(S)=\max\left\{v'(S), \max_{m_1\in [1:l-1]}\left\{v\left(\bigcup_{m=1}^{m_1} S_m \right)+v\left(\bigcup_{m=m_1+1}^l S_m\right)\right\}\right\},
    $$
    where $S=\bigcup_{m=1}^l S_m$, and $l$ is the size of the consecutive partition with each partition being largest.
\end{theorem}

Since both $v'(S)$ and $v(S_m)$ for consecutive subset $S_m\subseteq S$ can be obtained by the LLI algorithm, our \emph{Recursive LLI} algorithm for solving SDP (\ref{model2}) details as follows.

~\
\begin{breakablealgorithm}
	\renewcommand{\algorithmicrequire}{\textbf{Input:}}
	\renewcommand{\algorithmicensure}{\textbf{Output:}}
    \caption{\textsc{Recursive LLI Algorithm}}
    \label{alg2}

	\begin{algorithmic}[1]
		\Require Five-member tuple $(b,k,a,u,f)_i$ for all nodes in $S^{[1:s]}$, initial pollution level $k_{S^1-1}d_{S^1-1}$
        \Ensure Discharge solution $X$

        \Function{SubValue}{$S,k_{S^1-1}d_{S^1-1}$}
            \State Divide $S$ into consecutive and disjoint subsets $S_1,\dots,S_l$ with sizes as large as possible
            \State Let $\mathcal{T}'_{S^{[1:s]}}:=\mathcal{T}_S\cup (b_i,k_i,a_i,u_i,\beta\cdot x)_{i\in S^{[1:s]}\setminus S}$, where $\beta> \max\{K_1^if'_i(a_i),i\in S\}$
            \State Obtain $X'$ by inputting $(k_{S^1-1}d_{S^1-1},\mathcal{T}'_{S^{[1:s]}})$ in the LLI algorithm
            \State Get $v'(S):=\sum_{i\in S}f_i(x'_i)$
            \For{$m_1=1$ to $l-1$}
                \State Let $Y=(y_i)_{i\in \bigcup_{m=1}^{m_1} S_m}:=$\Call{SubValue}{$\bigcup_{m=1}^{m_1} S_m,k_{S^1-1}d_{S^1-1}$}
                \State Let $Z=(z_i)_{i\in \bigcup_{m=1}^{m_1} S_m}:=$\Call{SubValue}{$\bigcup_{m=m_1+1}^l S_m,k_{S_{m_1+1}^1-1}d_{S_{m_1+1}^1-1}$}, 
                \State Let $v_1:=\sum_{i\in \bigcup_{m=1}^{m_1} S_m}f_i(y_i)$; Let $v_2:=\sum_{i\in \bigcup_{m=m_1+1}^l S_m}f_i(z_i)$
                \If{$v_1+v_2> v'(S)$}
                    \State Update $X'=Y\cup Z \cup (\theta_i)_{i\in \left[\left(\max S_{m_1}\right)+1:\left(\min S_{m_1+1}\right)-1\right]}$
                \EndIf
            \EndFor
            \State \textbf{return} $X'$
        \EndFunction
        \State   
        \State \textbf{return} $X:=$\Call{SubValue}{$S,k_{S^1-1}d_{S^1-1}$}
	\end{algorithmic}  
\end{breakablealgorithm}

~\

It should be noted that the optimal discharge solution returned from the Recursive LLI algorithm also solves discharge quantities of nodes in $S^{[1:s]}\setminus S$.
In order to distinguish the optimal solutions under different sub-coalitions, we denote the optimal solution of SSDP (\ref{model2}) for $S$ as $X^S:=(x_i^S)_{i\in S^{[1:s]}}$ in the discussion below. 

Moreover, our main results derived in the Section \ref{sec3}, such as the LLI algorithm and the Recursive LLI algorithm can be directly applied to other more general cases where some profit functions have non-continuous fixed breakpoints and cost.
For example, if a profit function has a breakpoint, then the first order derivative at the breakpoint can be replaced by its right derivative.
And since we postulated the basic sewage discharge level, adding fixed cost will not make any difference to the results, so will not alter our algorithm.
In addition, the residual rate $k_i$ can be larger than $1$ for some $i\in N$.

\begin{remark}\label{rem1}
    Actually, the Recursive LLI algorithm is hard to implement when a sub-coalition is too scattered, i.e., neither of two nodes in the sub-coalition are consecutive. In this case, the LLI algorithm will operate at least $2^{l-1}$ times, and the time complexity will be intolerable. However, by Lemma \ref{lem7} from Section \ref{sec41}, the number of iterations of the Recursive LLI algorithm can be significantly reduced. As in the algorithm, we partition $S$ into $[S_1;\dots;S_l]$, and improve the dynamic programming of solving $v(S)$ as
    \begin{equation}\label{eq2}
        v(S)=\max_{m_1\in [2:l]}\left\{v'(S),v(\bigcup_{m=1}^{m_1-1} S_m)+v'(\bigcup_{m=m_1}^{l} S_m)\right\}.
    \end{equation}
    The time complexity of solving DP (\ref{eq2}) is at most $\mathcal{O}\left(l^2 (\max_{m\in [1:l]}s_m)^3\right)$.
\end{remark}

So far, we have studied the SDP for the grand coalition or a sub-coalition. This used as the building block to develop the cooperative game---sewage discharge game (SDG).

\section{Sewage Discharge Game}\label{sec4}

Our main focus on the \emph{cooperative game defined on the sewage discharge problem}, is investigating whether all firms would like to form a grand coalition so the core is non-empty and how to allocate the profit generated by the grand coalition to ensure the coalition stability. 
Unfortunately, it is well known that rare cooperative games have non-empty core, that is why our problem is difficult.
In the following, we will first discuss how nodes cooperate in different coalitions, upon which we are able to introduce properties of SDG, and then we will examine how a coalition is formed, thus we will be able to show core non-emptiness of SDG and obtain core allocations.

Based on the analyses of the SDP in Section \ref{sec3}, we can now formally introduce the sewage discharge game.
To be specific, given an SDP$(b_0,N)$, we describe the SDG by a pair $(N,v)$, where $N$ is the grand coalition of all nodes, and $v(\cdot): 2^N\rightarrow \mathbb{R}$, defined in programming (\ref{model2}), is the characteristic function that maps each coalition $S \subseteq N$ to a value $v(S)$, representing the maximum total profit guaranteed through cooperation among all players in $S$.

\subsection{Positional Advantage and Coalition Formation in SDG}\label{sec41}

Through solving an SDP, we can obtain an optimal sewage discharge scheme for each node in $N$. However, the SDP is a global optimization problem of distributing discharging quantities from the view of a central authority, which is not enough to describe self-incentives of firms. Obviously some subset of firms may have incentive to form coalition and transfer sewage discharge quota, which will cause externality to firms outside the coalition, and vice versa. In the following section we will analyze the externality and cooperation in SDG in terms of transfer of sewage discharge quota. 

\subsubsection{Positional Advantage of Nodes}

From Section \ref{sec3.1}, it is clear that the myopic discharge solution $(\theta_i)_{i\in N}$ is the resulting decision of $N$ in absence of the central authority intervention. Meanwhile, due to Theorem \ref{thm1} and $c_i(\Theta)=d_i$ for all $i\in N$, $(\theta_i)_{i\in N}$ can be seen as an intervention free sewage discharge quota myopically selected by each node. The optimal sewage discharge scheme of coalition $N$ can be seen as redistribution of sewage discharge quota $(\theta_i)_{i\in N}$. For example, the central authority transfers a portion of sewage discharge quota $\delta$ of a node $i_1$ to its downstream node $i_2$, so $i_2$ can obtain additional residual discharge quota $K_{i_1}^{i_2} \delta$. 
If the final optimal discharging quantity of a node is more than its myopic discharging quantity, this means that a portion of its discharge quota is obtained from its upstream node. The process of transferring discharge quota is the embodiment of cooperation between nodes.

With this perspective, it is more intuitive to interpret the externality caused by nodes outside the coalition. Think about an inconsecutive coalition $S$, so there is an outside node in between, i.e., $\exists i\in S^{[1:s]}$ but $i\notin S$. If the third party wants to transfer discharge quota from upstream nodes to downstream nodes in $S$, it has to first give up, i.e., transfer enough discharge quota to outside nodes in between, so make sure the discharge quotas of these nodes reach their top discharge level $u$. 
This is because nodes in between benefiting from their positional advantage, have the opportunity to occupy the transferred discharge quota that moves from upstream to downstream. Particularly, outside nodes are certain to capitalize on this opportunity.

Generally speaking, the positional advantage of nodes results in negative externality. This makes a portion of discharge quotas belong to the coalition ``wasted" and transferred to nodes outside the coalition. If there is a node outside the coalition with discharging quantity larger than its myopic discharge solution, then we can call it a \emph{free-rider} in the coalition. Nevertheless, should we leave them as free-riders?
Interestingly, we claim that having the free-riding nodes join the coalition would bring more profits than the highest profits those nodes can generate on their own. This is formalized in Proposition \ref{prop3}.
\begin{proposition}\label{prop3}
    Given an SDG $(N,v)$, for an inconsecutive coalition $S\subset N$ such that $S$ can be divided into two disjoint sub-coalitions $S_1$, $S_2$ with $S_1\prec S_2$. If $v(S)> v(S_1)+v(S_2)$, then
    $$
    v(S)+\sum_{S_1\prec i\prec S_2}f_i(u_i)\leq v(S\cup [\max S_1+1:\min S_2-1]).
    $$
\end{proposition}

Proposition \ref{prop3} fits the intuition very well. When a coalition creates mutual benefits through the cooperation, i.e., $v(S)>v(S_1)+v(S_2)$, it might offer an opportunity for outside nodes in between to free ride. Then some nodes (i.e. nodes in $[\max S_1+1:\min S_2-1]$) can discharge more since its upstream reduces sewage discharging quantity. Similar to the scenario discussed by \cite{sigman2002international}, an inconsecutive coalition does not lead to the full benefits of pollution control. Despite the side effects of the free-ride opportunities, coalitions that create mutual benefits often enjoy a sustained advantage \citep{kania2011collective}. When a coalition enhance the welfare of its members by creating mutual benefits, it also gains a greater social influence, which attracts more entities to join the coalition. Moreover, Proposition \ref{prop3} shows that bringing free-riders to coalition can further increase total benefits, which gives opportunity to improve welfare of each member. Now let us study how a coalition is formed in SDG.

\subsubsection{Mechanism for Coalition Formation}
In this section, we will particularly examine how a coalition is formed. For a coalition $S$, there will be $|S|!$ possible sequences that the nodes join the coalition, and after a new node joins how can we measure the discharge quota transferred among the nodes? In this paper, we introduce a measurement method to capture the discharge quota transferred between nodes, within or not within the coalition. This is based on a coalition formation mechanism as follows: For a coalition $S$, let the most upstream and the second-most upstream nodes in $S$ generate a subcoalition $\{S^1,S^2\}$ in advance, and then other nodes join in this subcoalition in sequence ordered from upstream to downstream, until all nodes in $S$ have joined this sub-coalition, i.e., the order of coalition formation is: 
$$\{S^1:S^2\}\rightarrow \{S^1,S^2,S^3\}\rightarrow \{S^1,S^2,S^3,S^4\}\rightarrow \cdots \rightarrow S\setminus S^s \rightarrow S$$ 
This coalition formation mechanism ensures the discharge quota is only transferred from upstream to downstream, and makes the measurement of the transferred volume more intuitive.

To formally axiomatize the measurement method, we set an auxiliary programming as in (\ref{model3}) below. Let $w(S,(x'_i)_{i\in R})$ be the maximum benefits generated by $S$ when the discharging quantities of nodes in $R$ are given as $x'_i$, where $S^{[1:s]}\setminus S\subseteq R\subseteq S^{[1:s]}$. We mathematically formulate $w(S,(x'_i)_{i\in R})$ as follows:
\begin{align}\label{model3}
    w(S,(x'_i)_{i\in R})=\max \quad & \sum_{i\in S}f_i(x_i)  & & \\
    s.t.\quad & \sum_{j=S^1}^i K_j^{i}x_j+ K_{S^1-1}^{i}d_{S^1-1} \leq b_i, & & \forall i\in S^{[1:s]}, \notag \\
    & a_i\leq x_i \leq u_i,& & \forall i\in S\setminus R, \notag \\
    & x_i=x'_i, & & \forall i\in R. \notag 
\end{align}
Since $(x_i)_{i\in R}$ is definite, all constraints are linear. $w(S,(x'_i)_{i\in R})$ can be regarded as a transformation of SDP (\ref{model1}). Obviously, its optimal solution can be obtained by the LLI algorithm. Let $(y_j(S,\cdot))_{j\in S^{[1:s]}}$ be the optimal solution of $w(S,\cdot)$.

In the following Definition \ref{def2} we use \emph{cooperation quantity} to measure the volume of discharge quota transferred between two nodes. The concept ``cooperation quantity" is very important for proofs of some Lemmas in this paper, but in fact, we have no requirement for its actual calculation. Let $x_i^S$ be the optimal discharging quantity of node $i$ in the SSDP$(d_{S^1-1},S,N)$.

\begin{definition}\label{def2}
    Given an SDG $(N,v)$, for coalition $S$ and nodes $i_1,i_2 \in N$, the volume of discharge quota transferred from $i_1$ to $i_2$ --- denoted by $\Delta_{i_1,i_2}^S$ --- is called as cooperation quantity between $i_1$ and $i_2$ in $S$. Let $\Delta_{i_1,i_2}^S=0$ if $i_2\leq i_1$ or $i_1\notin S$ or $i_2> S^s$. For $S^1\leq i_1< i_2\leq S^s$, we define
    \begin{numcases}{\Delta_{i_1,i_2}^S=}
        \Delta_{i_1,i_2}^{S\setminus S^s} & \emph{if } $x_{i_1}^S-x_{i_1}^{S\setminus S^s} \geq 0$  \notag \\
        \Delta_{i_1,i_2}^{S\setminus S^s} & \emph{if } $x_{i_2}^S-x_{i_2}^{S\setminus S^s} \leq 0$ \notag \\  \label{tag1} y_{i_1}(S,\hat{X}^S(i_2)\cup x_{i_2}^{S\setminus S^s}) - y_{i_1}(S,\hat{X}^S(i_2)\cup x_{i_2}^S) &  \emph{otherwise},
    \end{numcases}
    where $\hat{X}^S(i_2)=(x_j^S)_{j\in [S^1:i_2-1]\setminus J(i_2)}\cup (x_j^{S\setminus S^s})_{j\in [i_2+1:S^s]}$, with $J(i_2)=\{j\in S^{< i_2}: x_j^S< x_j^{S\setminus S^s}\}$.
\end{definition} 

After node $S^s$ joins coalition $S\setminus S^s$, for two nodes $i_1,i_2$ with $i_1<i_2$, we consider two cases: First is that $i_1$ obtains more discharge quota, i.e., $x_{i_1}^S\geq x_{i_1}^{S\setminus S^s}$. Second is that a portion of discharge quota of $i_2$ is transferred to other node, i.e., $x_{i_2}^S\leq x_{i_2}^{S\setminus S^s}$. Under both cases, as either the upstream $i_1$ discharges more or the downstream $i_2$ discharges less, there is no additional discharge quota transferred between $i_1$ and $i_2$ since $S^s$ joins $S\setminus S^s$, so $\Delta_{i_1,i_2}^S=\Delta_{i_1,i_2}^{S\setminus S^s}$.

While if neither the above two cases, after node $S^s$ joins coalition $S\setminus S^s$, an additional discharge quota $x_{i_2}^S - x_{i_2}^{S\setminus S^s}$ of node $i_2$ is transferred from some nodes in $J(i_2)=\{j\in S^{< i_2}: x_j^S< x_j^{S\setminus S^s}\}$. In light of the view positional advantage, node $i_2$ would obtain the additional discharge quota from $J(i_2)$ only when the discharging quantity of each node $j$ in $[S^1:i_2-1]\setminus J(i_2)$ reaches $x_j^S$. Thus, with the programming $w(S,\cdot)$, we can formulate the cooperation quantity between nodes $i_1$ and $i_2$ as expression (\ref{tag1}).
An example to to show how cooperation quantity measure the volume of discharge quota transferred in a coalition is provided in the Appendix \ref{appendix:examp}.

For coalition along river with line-positioning structure, compared with the constructive solution given by the algorithms in Section \ref{sec3}, sewage discharge quota transfer is more conducive to analyzing properties of cooperation.
Moreover, we claim that the coalition formation mechanism ensures that two nodes in a coalition will transfer discharge quota only once, as in Lemma \ref{lem6}.

\begin{lemma}\label{lem6}
    Given an SDG $(N,v)$, for a coalition $S$ and two nodes $i_1\in S$, $i_2\in [i_1+1:S^s]$, if  $\Delta_{i_1,i_2}^{S\setminus S^s}> 0$, then there is $\Delta_{i_1,i_2}^S = \Delta_{i_1,i_2}^{S\setminus S^s}$.
\end{lemma}

Furthermore, besides the free-riding behavior of the nodes outside a coalition, due to their positional advantage, Lemma \ref{lem6} also implies such free-riding is not easy to stop. Specifically, as shown in Lemma \ref{lem7}, for a node $i_2$ outside the coalition $S$, if $i_2$ takes a free-ride in $S$, i.e., there is a node $i_1\in S$ with $\Delta_{i_1,i_2}^S > 0$, then in a new coalition $S\cup j$ with downstream node $j\succ S$ joining, $i_2$ can continue its free-riding action, i.e. $\Delta_{i_1,i_2}^{S\cup j}=\Delta_{i_1,i_2}^S$. 

\begin{lemma}\label{lem7}
    Given an SDG $(N,v)$, for an inconsecutive coalition $S$ and a node $i\in S^{[1:s]}\setminus S$. If $v(S)>v(S^{<i})+v(S^{>i})$, then there is $v(S\cup j)> v(S^{<i})+v(S^{>i}\cup j)$ for all $j\succ S$.
\end{lemma}

According to Lemma \ref{lem7} and our coalition formation mechanism, we can conclude that the coalitions formed through this mechanism may be all ``stable". This stability arises from the fact that the joining of new downstream nodes does not disrupt the existing cooperation between the nodes within the coalition. Thus, we can confidently assert that the grand coalition $N$ is a stable coalition and that the SDG has non-empty core. We will show this in Section \ref{sec42}.

\subsection{Core Non-emptiness and Core Allocations for SDG}\label{sec42}

In cooperative game theory, core is an important concept related to stability of cooperation. A game with non-empty core means that the grand coalition is stable so no one has incentive to deviate and form subcoalition(see more descriptions in Section \ref{sec2.2}). However, to block each node deviating from the grand coalition, core allocation has exponential coalition stability constraints. Due to complexity of constraints, stable grand coalition is scarce in cooperative games. Getting the core non-emptiness of SDG is significantly important.

Consider a special case, all nodes in the grand coalition have a sufficient large sewage discharge upper bound, i.e., $u_i> b_i-K_1^i b_0$ for all $i\in N$. In this case, there is no node can take a free-ride. The value of any inconsecutive coalition is equal to the sum of values of its minimum number of consecutive disjoint subcoalitions. Therefore core allocation is easy to be solved from $(n+1)n/2-1$ coalition stability constraints. This special case can be found in existing works, i.e., in \citep{demange2004on}, which supposes that only consecutive coalition can be formed.
    
Nevertheless, the sewage discharge upper bound constraint of each node makes it possible for some nodes to take a free-ride in a coalition, and therefore inconsecutive coalition can be formed. This significantly increases the complexity of verifying and solving the core allocations. To investigate whether the core of SDG is non-empty, we also need Lemma \ref{lem8}.

\begin{lemma}\label{lem8}
    Given an SDG $(N,v)$, for two coalitions $S_1$ and $S_2$ with $S_1\prec S_2$, we have:
    $$
        v(S_1\cup S_2)\leq v(S_1)+\sum_{i\in S_2}f_i(u_i).
    $$
\end{lemma}

We would use Proposition \ref{prop3} and Lemma \ref{lem8} to show the non-emptiness of the core of SDG. This is built on the Bondareva-Shapley theorem \citep{1963Some,shapley1969market}, which shows a sufficient and necessary condition of the core non-emptiness.

\begin{remark}{\emph{(Bondareva-Shapley Theorem)}}
    A cooperative game $(N,v)$ with transferable utility has a nonempty core if, and only if, it is balanced. That is, the inequality $\sum_{S\subseteq N}\mu_Sv(S)\leq v(N)$ holds for any weight vector $(\mu_S)_{S\subseteq N} : \mu_S\in [0,1]$ such that $\sum_{S\ni i:S\subseteq N}\mu_{S}=1$ for each $i\in N$.
\end{remark}

\begin{theorem}\label{thm3}
   SDG $(N,v)$ is balanced, so its core is non-empty. 
\end{theorem}

The non-emptiness of the core of SDG could bring managerial insights. 
For firms along a linear river, non-empty core can induce them to make cooperation and form a stable grand coalition by properly allocating profit, so as to increase their wealth. 
For local governments, although they can only manage firms under their own jurisdiction, non-empty core can induce cooperation across regions, so to promote regional economic development without aggravating environmental pollution. 
Putting this to a bigger picture, for the entire society, economic production pollutes the environment initially. However, advanced economy would improve the environment sometime later \citep{grossman1995economic}. 
While non-empty core can induce cooperation among the entire society, so to advance the economic and reduce environmental pollution at the initial production stage, and sequentially achieve economic sustainable development.

Although we have shown that core allocations of SDG definitely exists, directly solving the core by programming (\ref{core}) is not easy. Fortunately, we find that the downstream incremental allocation proposed in \cite{ambec2002sharing} can also be a core allocation in the SDG, and this also supports our proof of the Theorem \ref{thm3}.

\begin{lemma}\label{lem9}
    Given an SDG $(N,v)$, there is a core allocation $(\alpha_i)_{i\in N}$ as
    $$
    \alpha_1 = v(\{1\}), ~ \alpha_i = v([1:i])-v([1:i-1]), ~ \text{ for all } ~ i\in [2:i].
    $$
\end{lemma}

So far, we have identified the core non-emptiness for SDG---rare for a general cooperative game---potentially because of the linear structure of the SDG, i.e., all player positioned a long a line. One interesting question we want to explore next is: can we extend the game to a more general framework and identify sufficient conditions for line-structured games to have non-empty core? 

\section{Games with Directional-convexity}\label{sec5}
So far, we have identified the core non-emptiness for SDG---rare for a general cooperative game---potentially because of the linear structure of the SDG, i.e., all player positioned a long a line. One interesting question we want to explore next is: can we extend the game to a more general framework?  In this section, we are interested in identifying more general sufficient conditions of core non-emptiness for games with players having heterogeneous advantages in cooperation. Specifically, heterogeneous advantages due to their positioning along a line.

\subsection{Directional-convexity}\label{sec5.1}
In the literature of cooperative game theory, one of the most well-known sufficient conditions for a game with transferable utility to have a non-empty core is that this game falls into the category of \emph{Convex Games} \citep[e.g.,][]{shapley1971cores}.
To be more specific, a cooperative game $(N,v)$ is a convex game if the inequality $v(S)+v(T)\leq v(S\cup T)+v(S\cap T)$ holds for any two coalitions $S,~T\subseteq N$.

While the definition of convex game looks general in testing the core non-emptiness of a TU-game, it is usually so strict that most of the existing games cannot satisfy.
For instance, although as shown in Theorem \ref{thm3} the core of an SDG is non-empty, we can see from Example \ref{exam2} that it is not a convex game.

\begin{example}\label{exam2}
Given an SDG defined on $(b_0,\mathcal{T}_N)$, where the grand coalition is $N=\{1,2,3\}$. The river pollution tolerance vector is $b=(3,7,10)$. The basic sewage discharge level is $a=(0,0,0)$. The maximum sewage discharge level is $u=(3,4,5)$. The initial pollution level is $b_0=0$. The residual rate vector is $k=(1,1,1)$. And the respective profit functions are $f_1(x_1)=20x_1-2x_1^2$, $f_2(x_2)=10x_2-x_2^2$ and $f(x_3)=20x_3-x_3^2$, for $i \in \{1,2,3\}$. We have the value for each coalition as:
$$
\begin{aligned}
    & v(\{1\})=42,~v(\{2\})=24,~v(\{3\})=51,~v(\{1,2\})=66,\\
    & ~v(\{1,3\})=96,~v(\{2,3\})=91,\text{ and }v(\{1,2,3\})=133.
\end{aligned}
$$
It is obvious that the given SDG is not a convex game. For example, considering two subsets $\{1,3\}$ and $\{2,3\}$ of $N$, we have that $v(\{1,3\})+v(\{2,3\}) = 96+91 > 133+51= v(\{1,2,3\})+v(\{3\})$. This violates the definition of convex game.
\end{example}

Noticing the fact that the SDG is non-convex while its core is non-empty, we are devoted to identifying new sufficient conditions of core non-emptiness for games such as SDG.
Below in Definition (\ref{def3}), let $\pi\in \Pi_N$ be a permutation (or an order) on the set of players, i.e., a one-to-one function from $N$ to $N$, we formally introduce this alternative sufficient condition as \emph{directional-convexity}.

\begin{definition}\label{def3}
    Given a TU-game $(N,v)$, its characteristic function $v(\cdot): 2^N\rightarrow \mathbb{R}$ is directional-convex if there is a permutation $\pi\in \Pi_N$ such that inequality $v(S)+v(T)\leq v(S\cup T)+v(S\cap T)$ holds for any two coalitions $S,~T\subseteq N$ such that $\min \left\{\pi(i):i\in T\setminus S\right\} > \max \left\{\pi(i):i\in S\cap T \right\}$ and $\min \left\{\pi(i):i\in S\setminus T \right\} > \min \left\{\pi(i):i\in S\cap T \right\}$. 
\end{definition}

Compared with convex game, game with directional-convex characteristic function is a relaxation of it. To be more specific, a cooperative game is a convex game is equivalent to the statement that its characteristic function is directional-convex for all permutation $\pi\in \Pi_N$.

The concept ``directional-convex" is based on the heterogeneous advantage of players, and the permutation $\pi$ on the set of players in Definition \ref{def3} is the embodiment of heterogeneous advantages of players. Intuitively, the directional-convex characteristic function means that the incentives for joining a coalition increase as the coalition grows in an order. This allows us to observe the emergence of the ``wild-goose-queue" effect. 

Take the SDG as an example, the line-positioning structure among players, captured by permutation $\pi(i)=i$ for all $i\in N$, confers heterogeneous advantages of players, e.g., the positional advantages of nodes described in Section \ref{sec41}. Due to this, when an upstream node joins a coalition, it serves as a catalyst, inspiring the downstream node to proactively join the coalition. This sequential coalition formation coincides with our mechanism proposed in Section \ref{sec41}.

In line with expectations, the SDG studied in this paper is a directional-convex game, whose characteristic function is directional-convex. 

\begin{theorem}\label{thm4}
    The characteristic function $v(S)$ of the SDG, computed by (\ref{model2}), is directional-convex.
\end{theorem}

The concept of directional-convex is not only limited to SDG.
In practice, there are other occasions where players have line-positioning structure in the cooperative game, such as in the supply chain management \citep{gopalakrishnan2021incentives}, electricity transmission \citep{gately1974sharing}, or even in the disease treatment \citep{duijzer2020core}. It is quite likely that their characteristic functions are directional-convex, and it is worthy checking their directional-convexity. The directional-convexity could serve as an alternative sufficient condition of ensuring core non-emptiness. The core non-emptiness in these operation management problems indicates the stable grand coalition. So cooperation in the grand coalition requires no intervention from outside.

\subsection{Core Non-emptiness for Games with Directional-convexity}\label{sec5.2}

From Section \ref{sec42}, as a special case of games with directional-convexity, the core non-emptiness of SDG is presented. We will now show that the core of a directional game is not empty by assigning core allocations. Additionally, we will compare this result with the case of a convex game.

For instance, consider a $3-$player game $(N,v)$ where $v(\cdot)$ is directional-convex for permutation $\pi(i)=i, \forall i\in N$. This implies that $v(\cdot)$ is superadditive and inequality $v(\{1,2\})+v(\{1,3\})\leq v(N)+v(\{1\})$ holds. Thus we have the relative positions of hyperplanes $\sum_{i\in S} \alpha_i=v(S)$ for all $S\subseteq N$. 
Notably, the difference with convex game is that $v(\{1,2\})+v(\{2,3\})\leq v(N)+v(\{2\})$ and $v(\{1,3\})+v(\{2,3\})\leq v(N)+v(\{3\})$ are not specified by directional-convexity, whereas the value of $v(\{2,3\})$ significantly influences whether these two inequalities hold.
In the following we would discuss the core of game $(N,v)$ for different values of $v(2,3)$.

\begin{figure}[H]
    \centering
    \subfigure{
    \includegraphics[width=9.5cm]{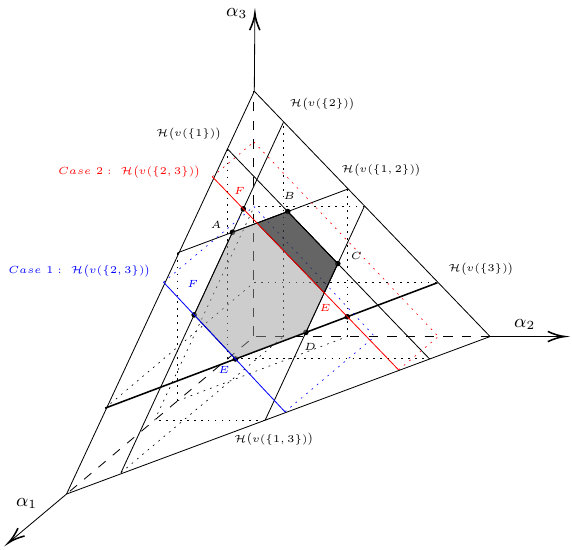}}
    \hfill
    \subfigure{
    \includegraphics[width=6.5cm]{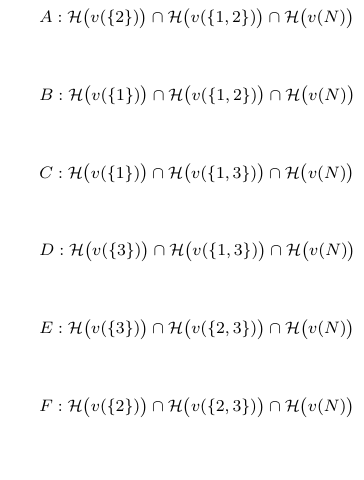}}   
    \caption{Core of a directional-convex game for different values of $v(2,3)$}
    \label{img2}
\end{figure}

Figure \ref{img2} exhibits the set of allocations $(\alpha_i)_{i\in N}\in \mathbb{R}_+^n$. Denote $\mathcal{H}\big(v(S)\big)$ as a hyperplane defined by $\sum_{i\in S}\alpha_i= v(S)$ for $S\subseteq N$. Points $A,B,C,D,E,F$ are intersections of hyperplanes. 
The blue line implies case $1$ that $v(\{2,3\})\leq v(N)-\max\{v(\{1,2\})-v(\{2\}),v(\{1,3\})-v(\{3\})\}$. In case $1$, the game $(N,v)$ becomes a convex game, the grey hexagon $ABCDEF$ represents the core. The red line implies case $2$ that $v(\{2,3\})>v(N)-\max\{v(\{1,2\})-v(\{2\}),v(\{1,3\})-v(\{3\})\}$. In case $2$, the black region represents the core. Consistent with expectations, as the value of $v(\{2,3\})$ increases, the core becomes narrower. Nevertheless, due to the directional-convexity, points $B$ and $C$ are invariably included as vertices of the core, regardless of the value of $v(\{2,3\})$.

In general, for a permutation $\pi\in \Pi_N$, we denote $S(\pi,j)=\{i\in N: \pi(i)\leq j\}$ for $j=1,\dots,n$ as the ``preorder" coalition. 
It is easy to find that a vertex of the core --- which is specified by the directional-convexity --- must be one of intersections $(\alpha_i)_{i\in N}$ of the hyperplanes $\mathcal{H}\big(v(S(\pi,j))\big)$, $j=1,\dots,n$, and the coordinate for each $i\in N$ is $\alpha_i=v\big(S(\pi,\pi(i))\big)-v\big(S(\pi,\pi(i)-1)\big)$. 

To obtain core vertices specified by the directional-convexity in a game $(N, v)$ with respect to a permutation $\pi$, we will now construct their corresponding permutations. Let $\psi\in\{0,1\}^n$ be an $n-$dimensional binary vector with $\psi_1=0$ and $\psi_n=0$. Define the ``rearrangements" of $\pi$ as $\pi_\psi$,
$$
\pi_\psi^{-1} (j) =
\begin{cases}
    \pi^{-1}\big(\max\{m< j:\psi_m=0 \}+1\big) & \text{if} ~ \psi_j=0 \\
    \pi^{-1}(j+1) & \text{if} ~ \psi_j=1
\end{cases}, \quad j=1,\dots,n;
$$
where $\pi_\psi^{-1}$ and $\pi^{-1}$ are the inverse functions of $\pi_\psi$ and $\pi$, respectively. In this way, each vector $\pi_\psi$ defines an intersection which corresponds to a core allocation, as stated in Theorem \ref{thm5}.

\begin{theorem}\label{thm5}
    Given a game $(N,v)$ having directional-convex $v(\cdot)$ with respect to permutation $\pi$, for each node $i\in N$, we have core allocation
    $$
    \alpha_i^\psi = v\left(S\big(\pi_\psi,\pi_\psi(i)\big)\right)-v\left(S\big(\pi_\psi,\pi_\psi(i)-1\big)\right).
    $$
\end{theorem}

While the relative positions of hyperplanes $\mathcal{H}\big(v(S)\big)$ within a game are not predetermined, the existence of directional-convexity ensures that the core of the game includes the convex hull of the vertices defined by $\pi_\psi$ for all $\psi$. For one of the vertices, the corresponding allocation of player $i$ can be seen as the marginal contribution it brings by joining the coalition in the designated order of $\pi_\psi$. This allows the central authority to flexibly utilize a convex combination of core allocations $(\alpha_i^\psi)_{i\in N}$, based on policy objectives, in order to distribute the overall benefits represented by $v(N)$ fairly among the players. Furthermore, applying the findings from this study to elucidate various practical phenomena is an intriguing avenue for further exploration.

\section{Conclusion}\label{sec6}

In this paper, we address the global optimization and the grand cooperation for sewage discharging decisions of firms. There are several firms located in a line along a river, which benefit from discharging sewage.
They can cooperate and form a coalition to generate more profits without exceeding the river pollution tolerances. Our purpose is to obtain the optimal sewage discharge scheme and profit allocations, so everyone will be willing to cooperate.

To solve the sewage discharge problem, we provide optimality conditions from two aspects---pollution level and marginal benefit. The conditions facilitate us to develop a greedy based algorithm to optimally solve SDP $v(N)$. 
While for sub-coalition $S$, the greedy strategy is insufficient. To obtain the optimal sewage discharge scheme of subcoalition $S$ in polynomial time, we construct a dynamic programming that builds upon the greedy for SDP. 
To show non-emptiness of the core of sewage discharge game, we discover several properties of the characteristic function and prove that the sewage discharge game is balanced.
In addition, inspired by the line-positioning of players in sewage discharge games, we propose a directional-convex game, as a relaxation of the convex game. 
To prove that the directional-convexity is a sufficient condition for cooperative games to have non-empty core, we capture $2^{n-2}$ vertices of the core space of a game with directional-convex characteristic function, and different convex combinations of these vertices can be used for a fairer profit allocation according to fit different destinations.

For future studies, there are still a lot interesting and worthy research questions. For example, how does cooperation work when multiple coalitions are allowed to exist simultaneously? Furthermore, what if the river structure is a sink tree or root tree? The results of our work can provide inspiration for addressing such questions.

\bibliographystyle{apalike}
\bibliography{ref}

\ECSwitch

\ECHead{Electronic Companion}

\section{Technical Proofs}\label{appendix:proof}

\vspace{1.5em}

\subsection{Proof of Theorem \ref{thm1}}

\proof{\scshape{Proof}.}
The given $X^*=(x_i^*)_{i\in N}$ is an optimal solution, and there is $\sum_{j=1}^i K_j^i x_j^* + K_1^i b_0\leq b_j$ for all $i\in N$. We define an error vector $(\varepsilon_i)_{i\in N}\in \mathbb{R}^n$ which is equal to $X^*- \Theta$ for $i\in N$, and for each $i\in [1:n]$, since the river pollution tolerance is $b_i$, there is 
$$
    c_i(X^*)=\sum_{j=1}^i K_j^i x_i^* = \sum_{j=1}^i K_j^i (\theta_j + \varepsilon_j) = d_i + \sum_{j=1}^i K_j^i \varepsilon_j \leq b_i - K_1^i b_0.
$$ 
And we have that
$$
    \sum_{j=1}^i K_j^i \varepsilon_j \leq b_i - K_1^i b_0 -d_i = b_i - K_1^i b_0 - \min\big\{\min_{j\in [1:i]}\{\sum_{l=j}^i u_j+k_{j-1}d_{j-1}\},b_i\big\}\leq - K_1^i b_0 \leq 0.
$$

Thus, we can get $c_i(X^*)=d_i + \sum_{j=1}^i K_j^i \varepsilon_j \leq d_i$. And if $\sum_{j=1}^n K_j^n \varepsilon_j < 0$, then we can find a vector $(\epsilon_i)_{i\in N}$ such that $(\epsilon_i\geq 0)_{i\in N}$ and $\sum_{j=1}^n K_j^n \epsilon_j + \sum_{j=1}^n K_j^n \varepsilon_j=0$. This implies $(x_i^*+\epsilon_i)_{i\in N}$ is a feasible solution which contradicted with that $X^*$ is the optimal solution, since all profit functions $f$ are monotonically increasing. Thus, there is $\sum_{j=1}^n K_j^n \varepsilon_j = 0$ and $c_i(X^*)=d_n$.

In fact, we can find that not only for the optimal solution $X^*$, there are $c_i(X)\leq d_i$, $\forall i\in N$ for any feasible discharge solution $X=(x_i)_{i\in N}$ of the SDP$(b_0,N)$.

\qed
\endproof

\subsection{Proof of Lemma \ref{lem1}}

\proof{\scshape{Proof}.}
Denote the revenue under the optimal discharge scheme $X^*=(x_i^*)_{i\in N}$ as $v_{max}$.
In the first case, for two adjacent nodes $i$ and $i+1$ with $i\in [1:n-1]$, if $x_i^*> a_i$ and $x_{i+1}^*< u_{i+1}$, then we can reduce the discharging quantity of $i$ by a small positive $\delta$ and increase the discharging quantity of $i+1$ by $k_i\delta$, and the resulting discharge scheme $X'=\{x_1,\dots,x_i-\delta,x_{i+1}+k_i\delta,\dots,x_n\}$ is also a feasible solution, the revenue under $X'$ is $v'$. The total change of the profit is
$$
\begin{aligned}
    \Delta_v=v'-v_{max} &= [f_{i+1}(x_{i+1}^*+k_i\delta)-f_{i+1}(x_{i+1}^*)]+[f_i(x_i^*-\delta)-f_i(x_i^*)] \\
    &\approx \delta[k_if'_{i+1}(x_{i+1}^*)-f'_i(x_i^*)]> 0.
\end{aligned}
$$
where the infinitesimal terms of Tylor expansion are omitted for simplification. $\Delta_v> 0$ contradicted with the optimality of $X^*$. Therefore, we have either $x_i^*=a_i$ or $x_{i+1}^*=u_{i+1}$ when inequality $f'_i(x_i^*)<k_if'_{i+1}(x_{i+1}^*)$ for $i\in [1:n-1]$.

In the second case, the proof process is similar to the first case. If $c_i(X^*)< d_i$ and $x_i^*< u_i$ and $x_{i+1}^*>a_{i+1}$, then we can increase the discharging quantity of $i$ by a small positive $\delta$ and reduce the discharging quantity of $i+1$ by $k_i\delta$. The total change of the profit is $\Delta_v\approx \delta[f'_i(x_i^*)-k_if'_{i+1}(x_{i+1}^*)]> 0$. Therefore, there is $c_i(X^*)= b_i$ or $x_i^*= u_i$ or $x_{i+1}^*=a_{i+1}$ when inequality $f'_i(x_i^*)>k_if'_{i+1}(x_{i+1}^*)$ for $i\in [1:n-1]$.

\qed
\endproof

\subsection{Proof of Lemma \ref{lem2}}

\proof{\scshape{Proof}.}
Divide the optimal solution into two parts: $X_{[1:j]}=(x_i^*)_{i\in [1:j]}$ and $X_{[j+1:n]}=(x_i^*)_{i\in [j+1:n]}$. We suppose there is an optimal solution for problem (\ref{model1}) to $[1:j]$ such as $X_{[1:j]}+\varepsilon$. $\pmb{\varepsilon}$ cannot be a positive vector since $c_j(X^*)=d_j$. Otherwise, if $\varepsilon > 0$, then $c_j(X_{[1:j]}+\varepsilon)>d_j$ which contradicted with the definition of $c_j(\cdot)$. That means the only way to obtain $v([1:j])$ is by setting $X_{[1:j]}$ as a feasible solution of SDP$(b_0,[1:j])$, and $\varepsilon$ has to meet
$$
(K_1^{j-1},K_2^{j-1},\dots,k_{j-1},1)\cdot\varepsilon=0.
$$
It shows that $(X_{[1:j]}+\varepsilon)\cup X_{[j+1:n]}$ is a feasible solution of problem (\ref{model1}). Since $\sum_{i\in [1:j]}f_i(x_i+\varepsilon_i)\geq \sum_{i\in [1:j]}f_i(x_i)$, we can get $(X_{[1:j]}+\varepsilon)\cup X_{[j+1:n]}$ is better than $X^*$, and this contradicted with the optimality of $X^*$. So $X_{[1:j]}$ is the optimal solution for problem (\ref{model1}) to $[1:j]$. In the same way, we can get $X_{[j+1:n]}$ is also the optimal solution for problem (\ref{model1}) to $[j+1:n]$. Thus, there is $v(N)=v([1:j])+v([j+1:n])$ if there is a node $j\in N$ with $c_j(X^*)=d_j$.

\qed
\endproof

\subsection{Proof of Proposition \ref{prop2}}

\proof{\scshape{Proof}.}
By greedy strategy, we denote the solution of problem (\ref{model1}) as $X^{(G)}$. For the set $N=\{1,2\}$, assume the optimal solution is $\{x_1^{(G)}-\delta,x_2^{(G)}+k_1\delta\}$ in which $\delta$ is a nonnegative value. From the concavity of $f$, we have
$$
f_1(x_1^{(G)}-\delta)+f_2(x_2^{(G)}+k_1\delta)< f_1(x_1^{(G)})-f'_1(x_1^{(G)})\delta + f_2(x_2^{(G)}) + f'_2(x_2^{(G)})k_1\delta.
$$
It is obvious to get $\delta=0$ and which refers $X^{(G)}$ is the optimal solution when the universe set only has two nodes. We suppose for set $\{1,,\dots,n-1\}$ can get the optimal solution by greedy strategy. Then it is sufficient to prove the optimality of greedy strategy for $N=\{1,,\dots,n\}$.

We can regard the problem to $N=\{1,,\dots,n\}$ as a sub-problem to set $\{1,\dots,n-1\}$ and node $n$ with discharging quantity $x_n^{(G)}$, in which the sub-problem has an extra constraint 
$$
\sum_{i\in N\setminus n}K_i^{n-2}x_i + K_1^{n-2}b_0 \leq \frac{b_n}{k_{n-1}} - \frac{x_n^{(G)}}{k_{n-1}}.
$$
And by greedy strategy, denote the optimal solution of this sub-problem as $X':=\{x'_1,\dots,x'_{n-1}\}$. Since change discharging quantity of one node will not affect the order of AMB of other nodes, we get $X^{(G)}=X'\cup \{x_n^{(G)}\}$. Assume there is a node $i$, for which we can find a positive value $\delta$ to get a better solution 
$$
X^*:=\{x'_1,\dots,x'_i-\delta,\dots,x'_{n-1},x_n^{(G)}+K_i^{n-1}\delta\}.
$$

That refers $K_i^{n-1}f'_n(x_n^{(G)})> f'_i(x'_i)$. However, greedy strategy requires nodes increase discharging quantities with their AMBs maintain equality. So if $K_i^{n-1}f'_n(x_n^{(G)})> f'_i(x'_i)$, then from Lemma \ref{lem1} we have $x_i=a_i$ or $x_n^{(G)}=u_n$. This contradicted with the nonnegativity of $\delta$. Therefore, we get the optimality of the greedy strategy for $N=\{1,,\dots,n\}$. And the proposition has been established as well.

\qed
\endproof

\subsection{Proof of Lemma \ref{lem3}}

\proof{\scshape{Proof}.}
For the LLI algorithm, it is obvious that the number of iterations is equal to the number of times the discharging quantities of nodes in $I^*$ have increased. There are three conditions that limit the increase of discharging quantities of nodes in $I^*$:
\begin{itemize}
    \item The discharging quantity of a node $i\in I^*$ reaches its upper bound level $u_i$,
    \item The AMBs of nodes in $I^*$ reduce to the highest AMB of a node outside $I^*$,
    \item The post-discharging pollution level at a node $i>\min I^*$ reaches $d_j$.
\end{itemize}
For an SDP$(b_0,N)$, each node $i\in N$ can only be put into $I^*$ at most once. When a node is removed from $I^*$, the discharging quantity of this node is reached optimal. In the worst case, where each update only involves the addition or removal of a single node from $i^*$, the number of iterations is $2n$.
Therefore, we can ascertain that the number of iterations in the LLI algorothm is at most $\mathcal{O}(n)$.

In each iteration, we have to calculate $\sigma_1$,$\sigma_2$,$\sigma_3$ once each. Notably, since for each $i\in N$, $f'_i(x_i)$ has analytical solution, we can conclude that values of both $g_{i^*}(\xi/K_1^{i^*})$ and $g_{i^*}(f'_{j^*}(u_{j^*})/K_{i^*}^{j^*})$ can be obtained in $\mathcal{O}(1)$. This means computational complexity for both $\sigma_1$ and $\sigma_2$ is not more than $\mathcal{O}(n)$. However, the value of $\sigma_3$ is defined by a linear program. The time complexity to calculate $\sigma$ may more than $\mathcal{O}(n)$. In other words, the time complexity of each iteration depends on the computational complexity for $\sigma_3$. 

It is easy to find that the linear program $\sigma_3:=\max\left\{\Delta x_{i^*}:\sum_{i\in I^*}^{i\leq j}K_i^j \Delta x_i \leq b_j-c_j(X), ~ \forall j\geq i^*\right\}$ is equivalent to $\sigma_3=\max\{\Delta x_{i^*}: \Delta x_{i}\leq \delta_i, \forall i\in I^*\}$. The value of vector $(\delta_i)_{i\in I^*}$ can be calculated in at most $\mathcal{O}(n)$, that is, $\sigma_3$ can be obtained in at most $\mathcal{O}(n^2)$.

Therefore, the time complexity of the LLI algorithm is at most $\mathcal{O}(n^3)$.

\qed
\endproof

\subsection{Proof of Lemma \ref{lem4}}

\proof{\scshape{Proof}.}
We suppose there is a node $i\in S^{[1:s]}\setminus S$ whose discharging quantity under the optimal scheme satisfies $\theta_i< x_i^*< u_i$, so there is $x_i^*=b_i-k_{i-1}c_{i-1}(X^*)$. From the proof of lemma \ref{lem2}, we can get $(x_i^*)_{i\in S^{[1:i]}}$ is the optimal solution of SSDP$(d_{S^1-1},S^{<i},N)$. Since $x_i^*>\theta_i$, there must be an upstream node $l\in S$ and $l< i$ whose discharging quantity $x_l^*<\theta_l$. Consider a solution $X'(S^{[1:i]}):=\{x_{S^1}^*,\dots,x_l^*+\delta,\dots,x_{i-1}^*,x_i^*-K_l^{i-1}\delta\}$ in which $\delta$ is a small positive value, then the objective function value has
$$
\sum_{j\in S}^{j< i}f_j(x'_j) > f_l(x_l^*) + \sum_{\substack{j\not= l \\ j\in S}}^{j< i}f_j(x'_j) = \sum_{j\in S}^{j< i}f_j(x_j^*).
$$
That is a conclusion contrary to the optimality of $(x_i^*)_{i\in S^{[1:s]}}$. Thus, nodes in $i\in S^{[1:s]}\setminus S$ cannot discharge sewage quantity between $\theta_i$ and $u_i$. 

On the other hand, since $c_{i-1}(X)\leq d_{i-1}$ for any feasible discharge solution $X$ of (\ref{model2}), we can get $x_i^*\geq b_i-k_{i-1}c_{i-1}(X^*)\geq \theta_i$ for all $i\in S^{[1:s]}\setminus S$. The lemma has been proved.

\qed
\endproof

\subsection{Proof of Lemma \ref{lem5}}

\proof{\scshape{Proof}.}
For two arbitrary disjoint sets $S_1,S_2\subset N$, let $(x_{1,i}^*)_{i\in S_1^{[1:s_1]}}$ and $(x_{2,i}^*)_{i\in S_2^{[1:s_2]}}$ are the optimal solution of $v(S_1)$ and $v(S_2)$, respectively. Denote $S$ as $S_1\cup S_2$, the solution 
$$
X'=(x'_i)_{i\in S^{[1:s]}}:=(x_{1,i}^*)_{i\in S_1}\cup (x_{2,i}^*)_{i\in S_2} \cup (x'_i)_{i\in \{S_1^{[1:s_1]}\cup S_2^{[1:s_2]}\}\setminus S} \cup (\theta_i)_{i\in S^{[1:s]}\setminus \{S_1^{[1:s_1]}\cup S_2^{[1:s_2]}\}}.
$$ 
is a feasible solution of $v(S)$, in which $x'_i=\min \{u_i,b_i-k_{i-1}c_{i-1}(X')\}$ for $i\in \{S_1^{[1:s_1]}\cup S_2^{[1:s_2]}\}\setminus S$. Thus, there is 
$$
v(S)\geq \sum_{i\in S} f_i(x'_i) = \sum_{i\in S_1} f_i(x_{1,i}^*) + \sum_{i\in S_2} f_i(x_{2,i}^*) = v(S_1)+v(S_2).
$$
And the superadditivity of $v(\cdot)$ for sub-coalitions $S\subset N$ has been confirmed.

\qed
\endproof

\subsection{Proof of Theorem \ref{thm2}}

\proof{\scshape{Proof}.}
We represent all subsets of $S$ in the following way: let $M_j$ and $R_j$ be two disjoint sets with $M_j\cup R_j=S_j$ for all $j\in [1:l]$, and $M:=\bigcup_{j=1}^l M_j$, $R:=\bigcup_{j=1}^l R_j$. Denote the optimal solution of SSDP$(d_{\cdot},\cdot,N)$ is $X^*(\cdot)$. There must be an $M=\bigcup_{i=1}^j S_i$ and an $R=\bigcup_{i=j+1}^l S_i$ (i.e., for the case with $\bigcup_{i=1}^j R_i=\varnothing$ and $\bigcup_{i=j+1}^l S_i=\varnothing$) for $j\in [1:l-1]$. It is easy to get that
$$
\begin{aligned}
    v(S)= & \max\{v'(S),\max_{M,R}\{v(M)+v(R)\}\} \\
    \geq & \max\{v'(S), \max_{j\in [1:l-1]}\{v(\bigcup_{i=1}^j S_i)+v(\bigcup_{i=j+1}^l S_i)\}.
\end{aligned}
$$

On the other side, under the optimal discharge solution $X^*(M)$ and $X^*(R)$, we can find out
$$
j^*=\min\{j: c_i(X^*(M))=c_i(X^*(R))=d_i,i=\max S_j, j\in [1:l] \}.
$$

Let $\tilde{S}_{i_1}^{i_2}:=[\min S_{i_1}:\max S_{i_2}]$ be a consecutive set covered by $i_2-i_1$ disjoint sets, and $\hat{S}_{i_1}^{i_2}:=[S_{i_1};\dots;S_{i_2}]$ be a union set of $i_2-i_1$ disjoint sets. If $j^*=l$, then $(x_i^*(M))_{i\in S^{[1:s]}\setminus R}\cup (x_i^*(R))_{i\in R}$ is a feasible solution of $v'(S)$. Otherwise, if $j^*< l$, then $(x_i^*(M))_{i\in \tilde{M}_1^{j^*}\setminus \hat{R}_1^{j^*}}\cup (x_i^*(R))_{i\in \hat{R}_1^{j^*}}$ is a feasible solution of $v(\hat{S}_1^{j^*})$, and $(x_i^*(M))_{i\in \tilde{M}_{j^*+1}^l\setminus \hat{R}_{j^*+1}^l}\cup (x_i^*(R))_{i\in \hat{R}_{j^*+1}^l}$ is a feasible solution of $v(\hat{S}_{j^*+1}^l)$. Hence, For any $M$ and $R$, there is 
$$
v(M)+v(R)\leq \max\{v'(S), \max_{j\in [1:l-1]}\{v(\bigcup_{i=1}^j S_i)+v(\bigcup_{i=j+1}^l S_i)\}.
$$

Therefore, we get
$$
v(S)=\max\{v'(S), \max_{j\in [1:l-1]}\{v(\bigcup_{i=1}^j S_i)+v(\bigcup_{i=j+1}^l S_i)\}.
$$

\qed
\endproof

\subsection{Proof of Remark \ref{rem1}}

\proof{\scshape{Proof}.}
The time consumption of the Recursive LLI algorithm to solve $v(S)$ is
$$
T(S)=\sum_{i=1}^{l-1}(T(\bigcup_{j=1}^i S_j)+T(\bigcup_{j=i+1}^l S_j)) + \mathscr{T}(s),
$$
in which $\mathscr{T}(s)$ is the time consumption of the LLI algorithm under $s$ nodes, and we have that $\mathscr{T}(s)=\mathcal{O}(s^3)$. Since $S_i$ are all consecutive sets for $i\in [1:l]$, there is $T(S_i)=\mathscr{T}(s_i)$. Then we have
$$
    T(S)= \sum_{i=1}^l\sum_{j=i}^l \mathscr{C}(\sum_{q=i}^j s_q) \mathscr{T}(\sum_{q=i}^j s_q),
$$
in which $\mathscr{C}(\cdot)$ is the coefficient of $\mathscr{T}(\cdot)$. For any $i_1,i_2\in [1:l]$ with $i_1\leq i_2$, there is 
$$
\mathscr{C}(\sum_{q=i_1}^{i_2} s_q)=\sum_{j=1}^{i_1-1}\mathscr{C}(\sum_{q=j}^{i_2} s_q) + \sum_{j=i_2+1}^l\mathscr{C}(\sum_{q=i_1}^j s_q),
$$ 
and $\mathscr{C}(s)=1$.

When $s_{i_1}=s_{i_2}$ for any $i_1,i_2\in [1:l]$, the time consumption would be
$$
T(S)= \sum_{i=1}^l 2^{l-i} (i\times s_l)^3 + \sum_{i=1}^{l-2} 2^i (s_l)^3 > \mathcal{O}(2^{l-1}(s_l)^3).
$$

Combine with Lemma \ref{lem7}, we can get that
$$
\begin{aligned}
    v(S)= & \max_{m_1\in [2:l]}\left\{v'(S),v(\bigcup_{m=1}^{m_1-1} S_m)+v'(\bigcup_{m=m_1}^{l} S_m)\right\} \\
    = & \max_{m_1\in [2:l]}\left\{v'(S), \max_{m_2\in [2:m_1-1]}\left\{ v'(\bigcup_{m=1}^{m_1-1} S_m),  v(\bigcup_{m=1}^{m_2-1} S_m)+v'(\bigcup_{m=m_2}^{m_1-1} S_m)\right\} +v'(\bigcup_{m=m_1}^{l} S_m)\right\} \\
    & \cdots \\
    = & \max_{m_1\in [1:l]} \left\{ v'(\bigcup_{m=1}^{m_1} S_m) + \max_{m_2\in [m_1+1:l]} \left\{ v'(\bigcup_{m=m_1+1}^{m_2} S_m ) + \max_{m_3\in[m_2+1:l]} \{ v'(\bigcup_{m=m_2+1}^{m_3} S_m ) + \cdots\} \right\} \right\}.
\end{aligned}
$$

\begin{figure}[h]
    \centering   

    \includegraphics[width=15cm]{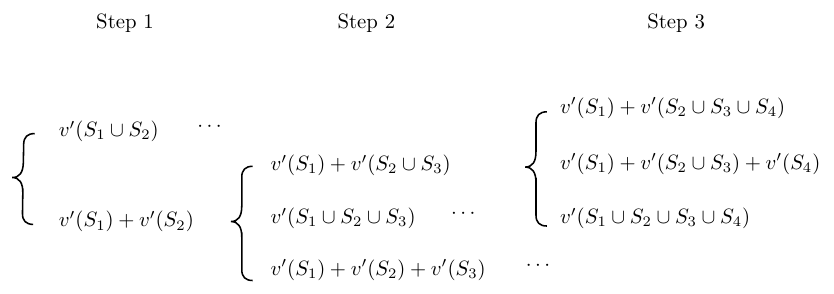}
    
    \caption{Process of solving $v(S)$}                               
    \label{img3}     
\end{figure}

For instance, if a coalition $S$ can be divided into four consecutive and disjoint sub-coalitions, i.e., $S_1,S_2,S_3,S_4$, then the step by-step process of solving $v(S)$ as in Figure \ref{img3}. At each step we choose a maximum value for the next step.

Thus, the optimal solution of $v(S)$ can be obtained in $l-1$ iterations by solving $v'(\bigcup_{m=m_1}^{m_2} S_m)$ for $m_1,m_2\in [1:l]$, and the time consumption for solving $v(S)$ would be reduced to
$$
T(S)=\sum_{m_1,m_2\in [1:l]} \mathscr{T}(\sum_{m=m_1}^{m_2} s_m)=\sum_{m_1,m_2\in [1:l]} (\sum_{m=m_1}^{m_2} s_m)^2 \leq \mathcal{O}\left(l^2 (\max_{m\in [1:l]}s_m)^3\right).
$$

\qed
\endproof

\subsection{Proof of Proposition \ref{prop3}}

\proof{\scshape{Proof}.}
Define $S_{1\cup2}=S_1\cup S_2\cup [S_1^{s_1}+1:S_2^1-1]$. If $v(S)>v(S_1)+v(S_2)$, then it is easy to get $x_i^S=u_i$ for all $i\in [S_1^{s_1}+1:S_2^1-1]$.
Thus $(x^S_i)_{i\in S}\cup (u_i)_{i\in [S_1^{s_1}+1:S_2^1-1]}$ is a feasible discharge scheme of $S_{1\cup 2}$, which implies $v(S_{1\cup 2})\geq v(S)+\sum_{S_1\prec i\prec S_2}f_i(u_i)$.

\qed
\endproof

\subsection{Proof of Lemma \ref{lem6}}

\proof{\scshape{Proof}.}
Given $S$ and two nodes $i\in S,j\in [i+1,S^s]$, By definition \ref{def2}, we can get that the cooperation quantity changes only when both $x_i^S < x_i^{S\setminus S^s}$ and $x_j^S > x_j^{S\setminus S^s}$ are true. Thus, it is sufficient to show that if $\Delta_{i,j}^{S\setminus S^s}> 0$, then $x_i^S < x_i^{S\setminus S^s}$ and $x_j^S> x_j^{S\setminus S^s}$ cannot hold simultaneously.

When $\Delta_{i,j}^{S\setminus S^s}> 0$, there are two cases:
\begin{itemize}
    \item Case $1$: $i$ and $j$ have equal marginal benefits such that:
    $$
    f'_i(x_i^{S\setminus S^s})=K_i^{j-1}f'_j(x_j^{S\setminus S^s})
    $$
    \item Case $2$: One of $i$ and $j$ has already reached its discharge bound such that:
    $$
    f'_i(x_i^{S\setminus S^s})< K_i^{j-1}f'_j(x_j^{S\setminus S^s})
    $$
    in which we have $x_i^{S\setminus S^s}=a_i$ or $x_j^{S\setminus S^s}=u_j$
\end{itemize}

Suppose $x_i^S < x_i^{S\setminus S^s}$ and $x_j^S> x_j^{S\setminus S^s}$ are both true, then case $2$ is impossible. In the case $1$, since the concavity of $f(\cdot)$, there is $K_i^{j-1}f'_j(x_j^S)< K_i^{j-1}f'_j(x_j^{S\setminus S^s}) = f'_i(x_i^{S\setminus S^s})< f'_i(x_i^S)$. Then we can find a better discharge solution $\{x_i^S+\xi,x_j^S-K_i^{j-1}\xi\}$ in which $\xi$ is a minimum positive value, and it contradicted with the optimality of $\{x_i^S,x_j^S\}$. Therefore, lemma is proved.

\qed
\endproof

\subsection{Proof of Lemma \ref{lem7}}

\proof{\scshape{Proof}.}
$v(S)>v(S^{<h})+v(S^{>h})$ implies that $x_i^S=u_i$ for all nodes $i$ such that $S^{<h}\prec i \prec S^{>h}$. Thus, some portion of sewage discharge quotas of these nodes must transferred from nodes in $S^{<h}$. Denote the set of these nodes as $I$, that is, for all $i\in I$ $\Delta_{i,h}^S> 0$. From lemma \ref{lem6}, we can make sure there is $\Delta_{i,h}^{S\cup l}> 0$ for each $i\in I$, that means $v(S)>v(S^{<h})+v(S^{>h}\cup l)$.

\qed
\endproof

\subsection{Proof of Lemma \ref{lem8}}

\proof{\scshape{Proof}.}
Let $l$ be a node in $N$ with $l\succ S$, we have
$$
\begin{aligned}
    v(S)+f_{l}(u_l)-v(S\cup l)=& \sum_{j\in S}f_j(x_j^S)+f_{l}(u_l)-\sum_{j\in S\cup l}f_j(x_j^{S\cup l}) \\
    =& \sum_{j\in S}[f_j(x_j^S)-f_j(x_j^{S\cup l})]+ f_{l}(u_l)-f_{l}(x_l^{S\cup l}).
\end{aligned}
$$
Based on the increasing of $f(\cdot)$, there is $f_{l}(u_l)\geq f_{l}(x_l^{S\cup l})$. In other side, since $\sum_{i\in S}K_i^{l-1}\Delta_{i,l}^{S\cup l} = x_l^{S\cup l}- \theta_l \geq 0$, it is obvious that $(x_i^{S\cup l})_{i\in S}$ is a feasible discharge solution of SSDP$(d_{S^1-1},S,N)$. Hence, $v(S)+f_l(u_l)-v(S\cup l)\geq 0$ is held. By analogy, we can deduce that $v(S\cup R)\leq v(S)+\sum_{i\in R}f_i(u_i)$ in which $R\subseteq N$ and $R\succ S$.

\qed
\endproof

\subsection{Proof of Theorem \ref{thm3}}

\proof{\scshape{Proof}.}
Let $\mathcal{N}=2^N\setminus \varnothing$ be the power set of $N$ and $\mathcal{N}_i:=\{S\in \mathcal{N}:S\ni i,i\in N\}$ be the set of coalitions containing the node $i$. 
    
The core of a SDG is non-empty if and only if $\sum_{S\in \mathcal{N}}\mu_Sv(S)\leq v(N)$ holds for any vector $(\mu_S)_{S\in \mathcal{N}} : \mu_S\in [0,1]$ with $\sum_{S\in \mathcal{N}_i}\mu_{S}=1: i\in N$. Since a decomposable coalition can be partitioned into several indecomposable coalitions, it is sufficient to prove $\sum_{S\in \mathcal{N}}\mu_Sv(S)\leq v(N)$ with postulate all coalitions are indecomposable.

To simplify the writing, we let $\mathcal{N}^{(m)}:=2^{[1:n-m]}\setminus \varnothing$ be the power set of $[1:n-m]$ and $\mathcal{N}_i^{(m)}:=\{S\in \mathcal{N}^{(m)}:S\ni i,i\in [1:n-m]\}$ in which $m\in [1:n-1]$. For any vector $(\mu_S)_{S\in \mathcal{N}} : \mu_S\in [0,1]$ with $\sum_{S\in \mathcal{N}_i}\mu_{S}=1: i\in N$, we can prove the theorem by split some coalitions to fill coalitions which contains $n$. There is
\begin{equation}\label{eq4}
    \sum_{S\in \mathcal{N}}\mu_Sv(S)= \sum_{S\in \mathcal{N}_n}\mu_Sv(S) + \sum_{m=1}^{n-1}\sum_{S\in \mathcal{N}_{n-m}^{(m)}}\mu_S v(S).
\end{equation}

Item $\sum_{S\in \mathcal{N}_{n-m}^{(m)}}\mu_S v(S)$ in equation (\ref{eq4}) would be split one by one in each iteration from $m=1$ to $n-2$. After an iteration, values of several disjoint coalitions can put together and form value of a bigger coalition by the superadditivity of $v(\cdot)$. We denote a vector $(\mu_S^{(j)})_{S\in \mathcal{N}_n}$ as the coefficient of coalition values after the $j^{\text{th}}$ iteration for $j\in [0:n-2]$, and $(\mu_S^{(0)}=\mu_S)_{S\in \mathcal{N}_n}$, $(\mu_S^{(j)}=0)_{S\in \mathcal{N}_n,S\subseteq [n-j+1:n]}$. For example, if $n=3$, then $v(\{1,2\})$ can be splited into $v(\{1\})$ and $f_2(u_2)$. Based on lemma \ref{lem8}, proposition \ref{prop3} and $v(\{2\})\leq f_2(u_2)$, there is
$$
\begin{aligned}
    & \mu_{\{1,2,3\}}v(\{1,2,3\}) + \mu_{\{2,3\}}v(\{2,3\}) + \mu_{\{1,3\}}v(\{1,3\}) \\
    & + \mu_{\{3\}}v(\{3\}) + \mu_{\{1,2\}}v(\{1,2\})+\mu_{\{2\}}v(\{2\}) \\
    \leq & \mu_{\{1,2,3\}}v(\{1,2,3\}) + \mu_{\{2,3\}}v(\{2,3\}) + \mu_{\{1,3\}}v(\{1,3\})+ \mu_{\{3\}}v(\{3\}) \\
    &  + (\mu_{\{1,2\}}-\min \{\mu_{\{1,2\}},\mu_{\{1,3\}}\}) v(\{1,2\})+\mu_{\{2\}}v(\{2\}) \\
    &  + \min \{\mu_{\{1,2\}},\mu_{\{1,3\}}\} (v(\{1\})+f_2(u_2)).
\end{aligned}
$$

Since $\mu_{\{1,2\}}+ \mu_{\{2\}}= \mu_{\{1,3\}} + \mu_{\{3\}}$, the right-hand side of above inequality is equal to 
$$
\begin{aligned}
    & \mu_{\{1,2,3\}}v(\{1,2,3\}) + (\mu_{\{1,3\}}-\min \{\mu_{\{1,2\}},\mu_{\{1,3\}}\}) (v(\{1,3\})+v(\{2\}))   \\ 
    & + (\mu_{\{1,2\}}-\min \{\mu_{\{1,2\}},\mu_{\{1,3\}}\}) (v(\{1,2\}) + v(\{3\}))  \\
    & + (\min \{\mu_{\{1,2\}},\mu_{\{1,3\}}\}+ \mu_{\{3\}}-\mu_{\{1,2\}})(v(\{3\})+v(\{2\})) \\
    & + \min \{\mu_{\{1,2\}},\mu_{\{1,3\}}\} (v(\{1,3\})+f_2(u_2)) + \mu_{\{2,3\}}v(\{2,3\}) \\
    \leq & ( \mu_{\{1,2\}} + \mu_{\{1,3\}}-2\min \{\mu_{\{1,2\}},\mu_{\{1,3\}}\} + \mu_{\{1,2,3\}}) v(\{1,2,3\}) \\
    & + (\min \{\mu_{\{1,2\}},\mu_{\{1,3\}}\}+ \mu_{\{3\}}-\mu_{\{1,2\}} + \mu_{\{2,3\}}) v(\{2,3\}) \\
    & + \min \{\mu_{\{1,2\}},\mu_{\{1,3\}}\} (v(\{1,3\})+f_2(u_2))  \\
    = & \mu_{\{1,2,3\}}^{(1)}v(\{1,2,3\}) + \mu_{\{2,3\}}^{(1)}v(\{2,3\}) + \mu_{\{1,3\}}^{(1)}(v(\{1,3\}) + f_2(u_2)).
\end{aligned}
$$

In the iteration, the sum of coefficients of coalitions containing $n$ remains $1$. After $j^{\text{th}}$ iteration, there is
$$
\begin{aligned}
    & \sum_{\substack{S\in \mathcal{N}_n \\ n-j-1\notin S}} \mu_S^{(j)} - \sum_{S\in \mathcal{N}_{n-j-1}^{(j+1)}}\mu_S \\
    = & \sum_{\substack{S\in \mathcal{N}_n \\ n-j-1\notin S}} \mu_S^{(j)} + \sum_{S\in \mathcal{N}_n\cap \mathcal{N}_{n-j-1}} \mu_S^{(j)} - \sum_{S\in \mathcal{N}_n\cap \mathcal{N}_{n-j-1}} \mu_S^{(j)} - \sum_{S\in \mathcal{N}_{n-j-1}^{(j+1)}}\mu_S \\
    = & \sum_{\substack{ S\in \mathcal{N}_n \\ S\not\subseteq [n-j+1:n] }} \mu_{S}^{(j)} - \sum_{i=0}^j\sum_{\substack{S\in \mathcal{N}_{n-i}^{(i)} \\ n-j-1\in S}} \mu_S - \sum_{S\in \mathcal{N}_{n-j-1}^{(j+1)}}\mu_S  \\
    = & \sum_{S\in \mathcal{N}_n} \mu_{S} - \sum_{S\in \mathcal{N}_{n-j-1}}\mu_S =0.
\end{aligned}
$$
for $j\in [0:n-2]$. In which $\sum_{S\in \mathcal{N}_n \cap \mathcal{N}_{n-j-1}}\mu_S^{(j)}=\sum_{i=0}^j\sum_{S\in \mathcal{N}_{n-i}^{(i)},n-j-1\in S}\mu_S$ can be explained as: in $\sum_{S\in \mathcal{N}_n}\mu_Sv(S) + \sum_{i=1}^{j+1}\sum_{S\in \mathcal{N}_{n-i}^{(i)}}\mu_S v(S)$,  $(v(S))_{n\in S}$ all come from $(v(S))_{S\in \mathcal{N}_n}$ and $(v(S))_{n-j-1\in S}$ all come from $(v(S))_{n-j-1\in S,S\in \bigcup_{i=0}^{j+1} \mathcal{N}_{n-i}^{(i)}}$.

Thus after the ${n-2}^{\text{th}}$ iterations, from equation (\ref{eq4}) we can get that 
$$
\begin{aligned}
    & \sum_{S\in \mathcal{N}_n}\mu_Sv(S) + \sum_{m=1}^{n-1}\sum_{S\in \mathcal{N}_{n-m}^{(m)}}\mu_S v(S) \\
    \leq & \mu_{\{1\}}v(\{1\})  + \sum_{\substack{S\in \mathcal{N}_n \\ S\not\subseteq [3:n]}}\mu_S^{(n-2)}\left(v(S)+\sum_{i\in N\setminus S}f_i(u_i)\right) \\ 
    = & \sum_{S\in \mathcal{N}_n\cap \mathcal{N}_1}\mu_S^{(n-2)}\left(v(S)+\sum_{i\in N\setminus S}f_i(u_i)\right) \\
    & + \sum_{\substack{S\in \mathcal{N}_n \\ 1\notin S \\ S\not\subseteq [3:n]}}\mu_S^{(n-2)}\left(v(\{1\}) + v(S)+\sum_{i\in [2:n]\setminus S}f_i(u_i)\right)\\
    \leq & \sum_{\substack{S\in \mathcal{N}_n \\ S\not\subseteq [3:n]}}\mu_S^{(n-2)} v(N)  =  v(N).
\end{aligned}
$$

This implies for the SDG $(N,v)$ is balanced. Consequently, the core of game $(N,v)$ is non-empty.

\qed
\endproof

\subsection{Proof of Lemma \ref{lem9}}

\proof{\scshape{Proof}.}
For the downstream incremental allocation, the budget balanced constraint of core is satisfied. Then we need to show coalition stability constraints.

For a coalition $S$, we partition it into minimum number of disjoint consecutive sub-coalitions, i.e. $S=[S_1;\cdots ; S_m]$. Then $[1:S^s]$ can be written as a union of $2m$ sets, i.e., $[T_1;S_1;\cdots;T_m;S_m]$. Hence, there is
$$
\sum_{i\in S}\alpha_i=\sum_{i=1}^m \left( v\big([1:\max S_i]\big) - v\big([1:\max T_i]\big) \right).
$$

Since the directional-convexity of $v(\cdot)$, we can get that
$$
\begin{aligned}
    & \sum_{i\in S}\alpha_i \\
    = & v([1:\max S_m])-v([1:\max T_m]) + v([T_1;S_1])-v(T_1) \\
    & + \sum_{i=2}^{m-1} \left( v([1:\max S_i]) - v([1:\max T_i]) \right) \\
    \geq & v([1:\max S_{m-1}]\cup S_m) - v([1:\max S_{m-1}])+ v([T_1;S_1])-v(T_1)  \\
    & + \sum_{i=2}^{m-1} \left(v([1:\max S_i]\cup \bigcup_{j=i+1}^m S_j) - v([1:\max T_i])\right. \\
    & \left. + v([1:\max S_i])- v([1:\max S_i]\cup \bigcup_{j=i+1}^m S_j)  \right) \\
    \geq & v([1:\max S_{m-1}]\cup S_m) - v([1:\max S_{m-1}])+ v([T_1;S_1])-v(T_1)  \\
    & + \sum_{i=2}^{m-1} \left(v([1:\max S_{i-1}]\cup \bigcup_{j=i}^m S_j) - v([1:\max S_{i-1}])\right. \\
    & \left. + v([1:\max S_i]) - v([1:\max S_i]\cup \bigcup_{j=i+1}^m S_j) \right) \\
    = & v(S).
\end{aligned}
$$

Therefore, the downstream incremental allocation is a core allocation of a game with directional-convexity.

\qed
\endproof

\subsection{Proof of Theorem \ref{thm4}}

\proof{\scshape{Proof}.}
In the SDG game $(N,v)$, we fix permutation $\pi(i)=i$ for $i\in N$. At the beginning, it is easy to find that the following two assertions are equivalent:
\begin{enumerate}[(I)]
    \item $v(\cdot)$ is directional-convex;
    \item $v(S\cup l) - v(S)\leq v(T\cup l) - v(T)$ for any two coalitions $S,T\subset N$ and a node $l\in N$ such that $T\setminus S \succ S $ and $l > S^1$.
\end{enumerate}
(I)$\Rightarrow$(II) is obvious. Let $U\subset N$ be a coalition such that $U^1=l$. Suppose (II) is true, then there is 
$$
\begin{aligned}
    & \begin{aligned}
        v(T)-v(S)\leq v(T\cup l)-v(S\cup l) \leq & v(T\cup \{U^1,U^2\}) - v(S\cup \{U^1,U^2\}) \\
        & \cdots \leq v(T\cup U) - v(S\cup U) \\
    \end{aligned} \\
    \Rightarrow & v(S\cup U) + v(T) \leq v\big( (S\cup U)\cap T \big) + v(S\cup U \cup T).
\end{aligned}
$$
where $S\cup U$ and $T$ corresponds to $S$ and $T$ in Definition \ref{def3}, respectively. That is, (II)$\Rightarrow$(I) is obtained.

Thus, given any two coalitions $S,T\subset N$ and a node $l\in N$ such that $T\setminus S \succ S $ and $l > S^1$, the proof of Theorem \ref{thm4} would be proceed from three aspects.

\begin{itemize}
    \item Firstly, when $l\succ T$.
\end{itemize}

In this case, we will discuss the directional-convexity in two situations. One is $v(S\cup l)=v(S)+v(l)$. It is easy to get the directional-convexity from super-additivity as
$$
v(T\cup l)-v(T)\geq v(l) = v(S\cup l)-v(S).
$$

Another is $v(S\cup l)>v(S)+v(l)$. In this situation, nodes between $S^s$ and $l$ can take free-riding such that $x_j^{S\cup l}=u_j$ for all $S\prec j<l$. $X'=(x_j^T)_{j\in R} \cup (x_j^{S\cup l})_{j\in S\cup l}$ is a feasible discharge scheme of $T\cup l$. Hence, there is
$$
\begin{aligned}
    v(T\cup l)-v(T) \geq & \sum_{j\in T\cup l}f_j(x'_j) - \sum_{i\in T}f_i(x_i^T) \\
    =  & \sum_{j\in S\cup l}f_j(x_j^{S\cup l})-\sum_{j\in S}f_j(x_j^T)  \\
    \geq & v(S\cup l) -v(S).
\end{aligned}     
$$

\begin{itemize}
    \item Secondly, when $S\prec l\prec T^t$.
\end{itemize}

We partition $R$ into two disjoint sub-coalitions $R_1$ and $R_2$ such that $R=R_1\cup R_2$ and $R_1\prec R_2$. Let $T_1:=S\cup R_1$. If $v(T_1)+v(R_2)=v(T)$, then we have
$$
\begin{aligned}
    v(S\cup l)-v(S) =& v(S\cup l) + v(R_2) - v(R_2)-v(S) \\
    \leq & v(S\cup R_1\cup l) +v(R_2) - v(S\cup R_1) -v(R_2) \\
    \leq & v(T_1\cup l\cup R_2) - v(T_1) - v(R_2) \\
    = & v(T\cup l) -v(T).
\end{aligned}
$$

And if $v(T_1)+v(R_2)< v(T)$, then we can obtain that the optimal discharge $x_l^T=u_l$. Obviously, $(x_i^T)_{i\in T\cup l}$ is a feasible discharge solution of $v(T\cup l)$. From proposition \ref{prop3} and lemma \ref{lem8}, there is
$$
v(S\cup l) - v(S) \leq f_l(u_l)\leq v(T\cup l) - v(T).
$$

\begin{itemize}
    \item At last, when $S^1< l < S^s$.
\end{itemize}

Given a node $j\in R$, $v(S\cup l\cup j)-v(S\cup l)\geq v(S\cup j)-v(S)$ is sufficient to the directional-convexity of $v(\cdot)$.

If $v(S\cup j)=v(S)+v(j)$, it is easy to get
$$
v(S\cup l \cup j)-v(S\cup l)\geq v(\{j\}) = v(S\cup j) - v(S).
$$
then we will discuss the directional-convexity of $v(\cdot)$ in two situations in the context of $v(S\cup j)>v(S)+v(j)$.

From lemma \ref{lem6}, we denote $S$ as an union set of a class of disjoint indecomposable subcoalitions, i.e., $S=\bigcup_{m_1=1}^m S_{m_1}$ with $S_1\prec\cdot\prec S_m$. There is a $m_1\in [1:m-1]$ with $S_{m_1}< l$, which makes $x_i^{S\cup l}\geq x_i^S$ for $i\in S_{m_1+1}\cup\dots\cup S_m$ and $x_i^{S\cup l}\leq x_i^S$ for $i\in S_1\cup\dots\cup S_{m_1}$. This is because nodes in $[S_1;\dots;S_{m_1}]$ can sell more discharge rights after $l$ joined $S$. That is to say $x_i^{S\cup l}\geq x_i^{\tilde{S}_{m_1}}$ for $i\in \tilde{S}_{m_1}$ where $\tilde{S}_{m_1}= S_{m_1+1}\cup\dots\cup S_m$. 

We construct a solution
$$
Y=(x_i^{S\cup l})_{i\in S\cup l\setminus \tilde{S}_{m_1}}\cup (x_i^{S\cup l}-(x_i^{\tilde{S}_{m_1}}-x^{S\cup j}))_{i\in \tilde{S}_{m_1}}\cup x_j^{S\cup j},
$$
in which $x^{S\cup j}\leq x_i^{\tilde{S}_{m_1}}$ for $i\in \tilde{S}_{m_1}$ since nodes in $\tilde{S}_{m_1}$ do not have cooperation relationship with nodes in $S\setminus \tilde{S}_{m_1}$. We can get that $Y$ is a feasible discharge solution of $v(S\cup j\cup l)$. Based on the concavity of $f(\cdot)$ and $(x_i^{S\cup j})_{i\in S\setminus \tilde{S}_{m_1}}$ is a feasible solution of $v(S\setminus \tilde{S}_{m_1})$, there is
$$
\begin{aligned}
    & v(S\cup  l\cup j)-v(S\cup l) \\
    \geq & \sum_{i\in S\cup l\cup j}f_i(y_i) - \sum_{i\in S\cup l}f_i(x_i^{S\cup l}) \\
    = & \sum_{i\in \tilde{S}_{m_1}} \left(f_i(x_i^{S\cup l}-(x_i^{\tilde{S}_{m_1}}-x^{S\cup j})) - f_i(x_i^{S\cup l})\right)  \\
    & + f_j(x_j^{S\cup j}) + \sum_{i\in S\setminus \tilde{S}_{m_1}}f_i(x_i^{S\cup j}) -\sum_{i\in S\setminus \tilde{S}_{m_1}}f_i(x_i^{S\cup j}) \\
    \geq & \sum_{i\in \tilde{S}_{m_1}} \left(f_i(x_i^{\tilde{S}_{m_1}}-(x_i^{\tilde{S}_{m_1}}-x^{S\cup j})) - f_i(x_i^{\tilde{S}_{m_1}})\right)  \\
    & + f_j(x_j^{S\cup j}) + \sum_{i\in S\setminus \tilde{S}_{m_1}}f_i(x_i^{S\cup j}) -v(S\setminus \tilde{S}_{m_1})\\
    = & \sum_{i\in \tilde{S}_{m_1}}f_i(x_i^{S\cup j})+ f_j(x_j^{S\cup j}) + \sum_{i\in S\setminus \tilde{S}_{m_1}}f_i(x_i^{S\cup j}) \\
    & - v(\tilde{S}_{m_1}) -v(S\setminus \tilde{S}_{m_1}) \\
    \geq & v(S\cup j) - v(S).
\end{aligned}
$$

With all three aspects above confirmed, we can get that the characteristic function $v(\cdot)$ of SDG $(N,v)$ is directional-convex.

\qed
\endproof

\subsection{Proof of Theorem \ref{thm5}}

\proof{\scshape{Proof}.}
Since $(\alpha_i^\psi)_{i\in N}$ is an intersection of the hyperplanes $\mathcal{H}\big(v(S(\pi_\psi,j))\big)$, $j=1,...,n$, it is clear that $\sum_{i\in N}\alpha_i^\psi = v(N)$ for each vector $\psi$, the budget balanced constraint of core is satisfied. Then we need to show coalition stability constraints.

Considering the $\pi_\psi$ can be regarded as an ordering in which players join the coalition, fix $\psi$, we redefine  the preorder coalition $S(\pi_\psi,j)$ by $\hat{N}_j$, and there are
$$
\hat{N}_0=\varnothing, \quad \hat{N}_j=\hat{N}_{j-1}\cup \pi_\psi(j),~ \forall j\in [1:n-1],\quad \hat{N}_n=N.
$$

For any coalition $S\subset N$, we will now show $\sum_{i\in S}\alpha_i^\psi \geq v(S)$. To simplify some notations, Let $\mathcal{P}(i)=\min\{j:i\in \hat{N}_j\}$, $\mathcal{P}^+(i)=\min\{\mathcal{P}(m):\mathcal{P}(m)>\mathcal{P}(i),m\in N\}$, $\mathcal{P}^-(i)=\max\{\mathcal{P}(m):\mathcal{P}(m)<\mathcal{P}(i),m\in S\}$ and $i^*=\arg\max \{\mathcal{P}(i): i\in S\}$. 

By the definition of $\hat{N}_j$, it is easy to get $\pi_\psi\big(\hat{N}_j\setminus \hat{N}_{j-1}\big)< \pi_\psi \big(N\setminus \hat{N}_j \big)$ for $j=1,\dots,n-2$. Thus for each $i\in S$, there must be a coalition $R_i\subset N$ with $\max{\pi_\psi(i):i\in\hat{N}_{\mathcal{P}^-(i)}}< \min\{\pi_\psi(i):i\in R_i\}$ such that $\hat{N}_{\mathcal{P}(i)-1}= \hat{N}_{\mathcal{P}^-(i)}\cup R_i$. That is,

\begin{align}\label{vertex}
    \sum_{i\in S} \alpha_i^\psi = & v(\hat{N}_{\mathcal{P}(i^*)}) - v(\hat{N}_{\mathcal{P}(i^*)-1}) + v(\hat{N}_{\mathcal{P}(S^1)}) - v(\hat{N}_{\mathcal{P}(S^1)-1}) \notag \\ 
    & + \sum_{\substack{i\in S \\ i\not = S^1,i^*}} \left( v(\hat{N}_{\mathcal{P}(i)}\cup (S\setminus \hat{N}_{\mathcal{P}^+(i)-1}))  - v(\hat{N}_{\mathcal{P}(i)-1}) \right. \notag \\ 
    & \left. + v(\hat{N}_{\mathcal{P}(i)}) - v(\hat{N}_{\mathcal{P}(i)}\cup (S\setminus \hat{N}_{\mathcal{P}^+(i)-1})).  \right)
\end{align}
    
Since the $v(\cdot)$ is directional-convex, we can get that
$$
\begin{aligned}
    (\ref{vertex})\geq & v(\hat{N}_{\mathcal{P}^-(i^*)}\cup i^*) - v(\hat{N}_{\mathcal{P}^-(i^*)}) + v(\hat{N}_{\mathcal{P}(S^1)}) - v(\hat{N}_{\mathcal{P}(S^1)-1}) \\
    & + \sum_{\substack{i\in S \\ i\not = S^1,i^*}} \left( v(\hat{N}_{\mathcal{P}^-(i)}\cup (S\setminus \hat{N}_{\mathcal{P}(i)-1}))  - v(\hat{N}_{\mathcal{P}^-(i)}) \right. \\
    & \left. + v(\hat{N}_{\mathcal{P}(i)}) - v(\hat{N}_{\mathcal{P}(i)}\cup (S\setminus \hat{N}_{\mathcal{P}^+(i)-1}))  \right) =v(S).
\end{aligned}
$$

Hence, for all $\psi$, allocations $(\alpha_i^\psi)_{i\in N}$ are core allocations of game $(N,v)$.

\qed

\section{An Example for an SDG}\label{appendix:examp}

To comprehend positional advantage and the coalition formation mechanism more accessibly, we consider a simple example. Given an SDG defined on $(b_0,\mathcal{T}_N)$, in which $N=[1:9]$, the river pollution tolerance vector is $b=(5,10,15,20,25,30,35,40,45)$, the basic sewage discharge level vector is $a=(0)_{[1:9]}$, the top sewage discharge level vector is $u_{[1:8]}=(5,6,5,6,5,6,6,6)$, the initial pollution level is $b_0=0$, the residual rate vector is $k=(1)_{[1:9]}$, and the respective profit functions are $f_1=3x$, $f_3=10x-x^2$, $f_5=20x-2x^2$, $f_7=20x-x^2$ and $f_9=40x-x^2$. 

Consider a coalition $S=\{1,3,5,7,9\}$, under the coalition formation mechanism, subcoalition $\{1,3\}$ formed in the first place, then node $5$ joins $\{1,3\}$ to form $\{1,3,5\}$, later node $7$ joins to form $\{1,3,5,7\}$, in the end node $9$ joins and coalition $S$ formed.
Before node $9$ joins $\{1,3,5,7\}$, it is not hard to get that the optimal discharge solutions of $v(\{1,3\})$, $v(\{1,3,5\})$ and $v(\{1,3,5,7\})$ are $(5,5,5)$, $(5,5,5,5,5)$ and $(5,5,3,6,4,6,6)$, respectively. 
In the subcoalition $\{1,3,5\}$, the central authority does not redistribute discharge quota since the myopic discharge solution is the optimal discharge scheme. However, after node $7$ joins $\{1,3,5\}$, the central authority can transfer discharge quota between nodes to obtain more coalition value, and we have that
$$
\begin{aligned}
    & \Delta_{3,4}^{\{1,3,5,7\}}=1, ~ \Delta_{3,6}^{\{1,3,5,7\}}=1/3, ~ \Delta_{5,6}^{\{1,3,5,7\}}=2/3, \\
    & \Delta_{3,7}^{\{1,3,5,7\}}=2/3, ~ \Delta_{5,7}^{\{1,3,5,7\}}=1/3.
\end{aligned}
$$

\begin{figure}[ht]
	\centering
    \subfigure[Case $1$: The top sewage discharge level of node $9$ is $u_9=6$]{
        \centering
        \includegraphics[width=12cm]{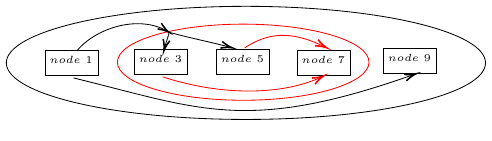}

        \label{img1a}
    }
    \subfigure[Case $2$: The top sewage discharge level of node $9$ is $u_9=11$]{
        \centering
        \includegraphics[width=12cm]{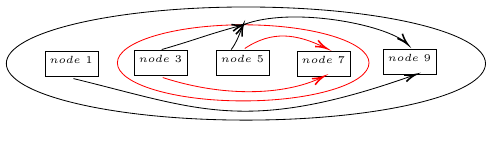}

        \label{img1b}
    }
    \caption{Transfer of sewage discharge quota}                               
    \label{img1}   
\end{figure}

This can be graphically illustrated in Figure \ref{img1}. The red arrows and red ellipse represent the transfers of discharge quota and a coalition partition which is taken free-riding, respectively. Accordingly, the black arrows represent newly generated transfers of discharge quota after node $9$ joins $\{1,3,5,7\}$, and black ellipses represent a coalition partition in $\{1,3,5,7,9\}$ and is taken free-riding.

After node $9$ joins $\{1,3,5,7\}$, we consider two cases. 
In the first case, the top sewage discharge level of node $9$ is $u_9=6$. Then the optimal discharge solution of $v(S)$ is $(5/4,6,7/2,6,17/4,6,6,6,6)$, and there are five newly generated discharge quota transfers, i.e.,
$$
\Delta_{1,2}^S=\Delta_{1,8}^S=\Delta_{1,9}^S=1, ~ \Delta_{1,3}^S=1/2, ~ \Delta_{1,5}^S=1/4.
$$
which can be graphically illustrated in Figure \ref{img1a}.

In another case, the the top sewage discharge level of node $9$ is $u_9=11$. Then the optimal discharge solution of $v(S)$ changes to $(0,6,1,6,3,6,6,6,11)$, and the newly generated discharge quota transfers as
$$
\Delta_{1,2}^S=\Delta_{1,8}^S=\Delta_{5,9}^S=1, ~ \Delta_{1,9}^S=3, ~ \Delta_{3,9}^S=2.
$$
which can be graphically illustrated in Figure \ref{img1b}. 

In both cases, existing discharge quota transfers are not affected by the newly joined downstream node, i.e., there are $\Delta_{3,4}^S=1$ $\Delta_{3,6}^S=1/3$, $\Delta_{5,6}^S=2/3$, $\Delta_{3,7}^S=2/3$ and $\Delta_{5,7}^S=1/3$ as well, which implies nodes $4$ and $6$ can continue to take a free-ride in coalition $S$.

\end{document}